\documentclass[review]{elsarticle}

\usepackage{lineno,hyperref}
\modulolinenumbers[5]

\usepackage{graphicx}
\usepackage{makeidx}  
\usepackage[english]{babel}
\usepackage{latexsym}
\usepackage{amssymb}
\usepackage{amsfonts}
\usepackage{amsmath}
\usepackage{graphicx}
\usepackage{xcolor}
\usepackage{multicol}
\usepackage{enumitem}
\usepackage{tikz}
\usepackage{mathdots}
\usepackage{diagbox}
\usepackage{adjustbox}
\usepackage{dashbox}
\usepackage{cleveref}
\usepackage{cancel}
\usepackage{color}
\usepackage{todonotes}
\usepackage{array}
\usepackage{multirow}
\usepackage{amssymb}
\usepackage{amsthm}
\usepackage{gensymb}
\usepackage{tabularx}
\usepackage{booktabs}
\usepackage{longtable}
\usepackage{geometry}
\usepackage{caption}
\usepackage{esvect}

\usetikzlibrary{fadings}
\usetikzlibrary{patterns}
\usetikzlibrary{shadows.blur}
\usetikzlibrary{shapes}
\usetikzlibrary{arrows,automata,positioning}
\tikzset{>=latex}

\usepackage{todonotes}

\graphicspath{ {./images/} }

\newcommand{\N}{\mathbb N}

\newcommand{\Z}{\mathbb Z}

\newcommand{\s}{\mathcal{S}}
\newtheorem{theorem}{Theorem}
\newtheorem{corollary}{Corollary}

\newtheorem{lemma}{Lemma}
\newtheorem{proposition}{Proposition}

\newtheorem{remark}{Remark}

\usepackage{tabularx,lipsum,environ,amsmath,amssymb}

\makeatletter

\journal{Theoretical Computer Science}

\begin{document}

\begin{frontmatter}

\title{Dynamical stability of threshold networks over undirected signed graphs}

\author[mymainaddress]{Eric Goles}
\author[mymainaddress]{Pedro Montealegre}
\author[mymainaddress]{Mart\'in R\'ios-Wilson}
\author[mysecondaryaddress,mythirdaddress]{Sylvain Sen\'e}

\address[mymainaddress]{Facultad de Ingenier\'ia y Ciencias, Universidad Adolfo Ib\'a\~nez.}
\address[mysecondaryaddress]{Universit{\'e} publique, Marseille, France}
\address[mythirdaddress]{Aix Marseille Univ, CNRS, LIS, Marseille, France}

\fntext[thanks]{This research was partially supported by ANID FONDECYT Postdoctorado 3220205 (M.R.-W.), ANID FONDECYT 1230599 (P.M.), BASAL funds for centers of excellence from ANID-Chile (E.G. and P.M.), STIC AmSud 22-STIC-02 CAMA project (E.G, M.R.-W., P.M. and S.S.), and European Union MSCA-SE-101131549 ACANCOS project (E.G., P.M. and S.S.).}

\begin{abstract}

In this paper, we explore the dynamics of threshold networks on undirected signed graphs. Much attention has been dedicated to understanding the convergence and long-term behavior of this model. Yet, an open question persists: How does the underlying graph structure impact network dynamics? Similar studies have been carried out for threshold networks and other types of Boolean networks, but the latter primarily focus on unsigned networks. Here, we address this question in the context of signed threshold networks.

We introduce the \emph{stability index} of a signed graph, related to the concepts of antibalance in signed graphs. Our index establishes a connection between the structure and the dynamics of signed threshold networks. We show that signed graphs having a negative stability index on every induced subgraph exhibit stable dynamics, i.e., the dynamics converge to fixed points regardless of their threshold parameters. Conversely, if at least one induced subgraph has a non-negative stability index, oscillations in long-term behavior may appear. Furthermore, we generalize the analysis to network dynamics under periodic update schemes. 

\end{abstract}

\begin{keyword}
 discrete dynamical systems \sep Boolean networks \sep threshold networks \sep stability index
\end{keyword}

\end{frontmatter}

\section{Introduction}

Originating from McCulloch and Pitts~\cite{mcculloch1943}, threshold networks were conceived as an initial model for the nervous system. These networks feature units interconnected by threshold functions, emulating neuron behavior~\cite{rosenblatt1958,minsky1969,hopfield1982}. In addition, Thomas and Kauffman applied similar principles to gene interaction modeling using Boolean functions~\cite{kauffman1969,thomas1973}. Since the end of the 1980\textit{s}, threshold functions have played a notable role in gene interaction models, with multiple studies exploring dynamics and resilience of cell cycle networks for instance~\cite{li2004,davidich2008,ruz2014}. In this paper, we study threshold networks where the interactions between the entities are either friendly or unfriendly. The model is defined by pairs $(G,w)$, where  $G = (V, E)$ is an undirected graph, with $V$ the set of entities (or vertices) and $E$ the set of interactions (or edges) between them. To each of these edges $e$ it is assigned a \emph{weight} $w(e)\in\{-1,+1\}$, representing the type of interaction between the endpoints of $e$,  which can be friendly (edge labeled by $+1$) or unfriendly (edge labeled by $-1$). Each node holds an internal state from the set~\(\{-1, +1\}\). This state evolves over discrete time-steps according to a deterministic local threshold transition function.  More precisely, a vertex adopts a $+1$ state (it is activated) if the weighted sum of the weights coming from its activated neighbors surpasses its predefined threshold; otherwise, the node adopts a $-1$ state.
  
Studying the dynamics of such networks requires to define when the entities execute their local function over time, namely an update scheme. The primary update scheme is the parallel one, where all nodes undergo synchronous updates. However, we are also interested in a broader spectrum of update schemes, with a particular emphasis on periodic update modes. The latter are defined by a finite sequence of sets $\mu = (A_1, \ldots, A_p)$, where each $A_i$ is a subset of $V$. During each time step, updates occur sequentially across sets, from the first to the last, with parallel updates within each set. A single entity may belong to multiple sets, which can make it change its state several times within a single time step. The question of the relations between the structure of interaction networks and their dynamic properties has been first addressed in~\cite{robert1980,goles1980,goles1982} and then developed from both combinatorial~\cite{noual2012,dns12,goles2015} and complexity standpoints~\cite{goles2015complexity,bridoux2022}; that of (a)synchronism sensitivity (the impact of update modes on network dynamics) which underlies is, for its part, more recent~\cite{salinas2008,aracena2013,noual2018,rios2021,donoso2024}.

In this context, this work focuses on the connections between signed threshold networks and their dynamics, according to the two aforementioned dimensions: firstly, the structural landscape defined by the structure of the graph; secondly, the dynamic evolution of the latter depending on various update modes.

This type of study already exists for unsigned threshold networks (where the edge weights are all equal to 1). In particular, Goles and Ruz~\cite{goles2015} defined, for each graph $G$, a parameter $\alpha(G)$ that depends on the number of nodes, edges, and self-loops in $G$, as well as on a parameter $p(G)$, which is the minimum number of edges that must be removed from $G$ to make it bipartite. The authors showed that the sign of $\alpha(G)$ is related to the asymptotic behavior of any threshold network that can be defined on $G$. More precisely, if $\alpha(G')<0$ for every induced subgraph of $G$, then every threshold network defined on $G$ admits only fixed points. Conversely, if $\alpha(G')\geq 0$ for any subgraph $G'$ of $G$, then it is possible to define a threshold network on $G$ that admits attractors other than fixed points.

In this article, we generalize the result of Goles and Ruz~\cite{goles2015} to signed graphs. To do so, for each signed graph $(G,w)$, we introduce the stability index $\s(G,w)$, which is defined similarly to $\alpha(G)$, but instead of considering the parameter $p(G)$, it considers a parameter $\rho(G,w)$, which is the minimum set of edges that must be removed from $G$ to obtain an \emph{antibalanced} graph. The property of being \emph{antibalanced} is a generalization of bipartite to signed graphs. Leveraging this stability index leads to two relevant and promising results for the study of signed threshold networks, analogous to those obtained in~\cite{goles2015}.

Furthermore, we extend the analysis of~\cite{goles2015} to the study of periodic updating schemes: we show that given a threshold network $\mathcal{T}$ defined on an undirected signed graph $(G,w)$ and a periodic update mode $\mu$, the negativity of the stability index of $(G,w)$ is a sufficient condition for the underlying dynamic system, to have only fixed points.

\subsection{Structure of the article}

In \Cref{sec:prelim} we present the main definitions and notation used in the paper. In \Cref{sec:stabind} we introduce the stability index of a graph and we present some preliminary results on the link between this parameter and the asymptotic behavior of the dynamics of threshold networks. In Section~\ref{sec:par}, we study the synchronous update scheme. In particular, we show that the negativity of the stability index is a sufficient condition for the network to admit only fixed points, and present results on what can happen when this index is not negative. In Section~\ref{sec:periodic}, we study the stability of the network under general periodic update schemes. We deduce from the previous section that the main theorem for synchronous update schemes can be extended to this setting. 
Finally, we explore stability for update schemes with constant block size. In particular, following the same approach than in \cite{goles2015}, we study all the graphs with positive self-loops in each node and with at most $4$ nodes. Interestingly, we show that the result stated in \cite{goles2015} for partitions of size at most $3$ does not hold for signed graphs. However, we provide a complete list of graphs of size $3$ and $4$ that are not stable and we show that if they are not induced subgraphs of the original network then, the latter is stable.

\section{Preliminaries}
\label{sec:prelim}

\paragraph{Signed graphs} A \emph{signed graph} is a pair $(G=(V,E), w)$, where $G$ is an undirected graph (possibly with self-loops) with a set of nodes $V$, a set of edges $E$ and where $w: E \to \{-1, 1\}$ is a \emph{sign assignment} for each edge in $G$. In addition, we denote by $n$ the number of nodes in $G$, i.e., $n = |V|$, by $m$ the number of edges (not counting self-loops) of $G$, and by $d^{+}$ (resp. $d^{-}$) the number of positive (resp. negative) self-loops of $(G,w)$.  Notice that $|E| = m + d^- + d^+$. 

In the following, the set of vertices $V$ of an $n$-node graph $G$ is identified with the set $[n] = \{1, \dots, n\}$. Furthermore, we denote $w_{ij}$ as the sign $w(\{i, j\})$, and the matrix $W = (w_{ij})$ is called the \emph{interaction matrix} of $(G, w)$. Given  a node $i$ of $G$, we define the \emph{neighborhood} of $i$, denoted by $N(i)$, as the set of nodes $\{j \in V\setminus \{i\}: \{i,j\} \in E\}$. The \emph{degree} of a node is the cardinality of its neighborhood. Given a signed graph $(G,w)$ and a subgraph $H$ of $G$, we abuse notation denote $(H,w)$ the signed subgraph $(H, w|_{E(H)})$. In the following each time that we mention a subgraph of a signed graph, we mean the signed subgraph.

\paragraph{Threshold networks} A \emph{threshold network} is a tuple $\mathcal{T} = (G,w, b)$, where $(G,w)$ is an $n$-node signed graph, $b \in \Z^{n}$ is a \emph{threshold vector}. Associated to a threshold network, we define a collection $f = (f_1, \ldots, f_n)$ of \emph{threshold local functions} 
$$f_i: \{-1,1\}^{n} \to \{-1,1\}$$ which are defined as:
\begin{equation}\label{eq:1}
  f_i(x) = \begin{cases}
    1 & \text{if } \sum_{j \in V} w_{ij} x_j - b_i > 0\text{,}\\
    x_i & \text{if } \sum_{j \in V} w_{ij} x_j - b_i = 0\text{,}\\
    -1 & \text{otherwise,}  
  \end{cases}
\end{equation}
where    $x \in \{-1,1\}^{n}$ is called a \emph{configuration of $\mathcal{T}$}.

We also define the \emph{global transition function} or simply \emph{the global rule} of the network as the map $F:Q^V \mapsto Q^V$ as $F(x)_i = f_i(x).$ The global transition function $F$  defines a dynamical system on the set of configurations. Given an initial condition $x^0 = x$, we define $x^t = F(x^{t-1})$ for all $t \geq 1$. The sequence of configurations $(x^t)_{t\geq 0}$ is called the \emph{orbit} of $x$. 

Observe that the second case in \Cref{eq:1} defines what to do in a tie case scenario, i.e., the case in which for some node $i$, the sum of the states of its neighbors is exactly $b_i$. For example,  the  majority rule, i.e., when $b_i = 0$, a tie-case scenario happens if for some node $i$ the number of neighbors in state $1$ is the same as that of neighbors in state $-1$. Of course, this can only happen if a node has even degree. When the majority rule is defined in this way, it is usually called \emph{stable}. In addition, other \emph{tie-breaking} functions can be defined. For instance, the majority rule is known in the literature as \emph{crazy spin} or \emph{unstable}, when a node changes its state in a tie-case scenario.  Observe that the stable majority can be defined with a positive self-loop, while the unstable majority is defined with a negative one. Unsurprisingly, the choice of the tie breaking rule has a great impact on the dynamics. We illustrate this with the example in Figure~\ref{fig:stabvsunstab}.

Since the number of possible configurations is finite, each orbit is \emph{eventually periodic}, i.e., there exists some $\tau, p \geq 0$ such that $x^{t+p} = x^t$ for all $t >\tau$. For a fixed orbit, the minimum time $t^*$ such that there exists $p$ satisfying $x^{t^*+p} = x^{t^*}$ is called the transient length of the orbit and  the sequence $x^{0},\hdots, x^{t^*}$ is called the \emph{transient} of the orbit.  A periodic orbit of period $p$ is called an \emph{attractor of period $p$}. In the case in which $p = 1$ the attractors are called \emph{fixed points}. The other attractors of a threshold network are called \emph{limit cycles}.  We  distinguish the limit-cycles of period $2$, which are called \emph{limit two-cycles}. A limit two-cycle in which each node changes its state, i.e., $x_i \neq F_\mu(x)_{i}$ for all $i \in V$, is called a  \emph{total limit cycle}. Figure~\ref{fig:twototal} depicts a total limit cycle and a limit two-cycle. We say that a threshold network is \emph{stable} if it only admits fixed points as attractors. Otherwise, we say that the network is \emph{unstable}.

\begin{figure}[t!]

\tikzset{every picture/.style={line width=0.75pt}} %set default line width to 0.75pt        

\begin{tikzpicture}[x=0.6pt,y=0.6pt,yscale=-1,xscale=1]
%uncomment if require: \path (0,452); %set diagram left start at 0, and has height of 452

%Shape: Square [id:dp3452731278307747] 
\draw  [fill={rgb, 255:red, 255; green, 255; blue, 255 }  ,fill opacity=1 ] (113,51) -- (205,51) -- (205,143) -- (113,143) -- cycle ;
%Shape: Circle [id:dp83452386078391] 
\draw  [fill={rgb, 255:red, 0; green, 0; blue, 0 }  ,fill opacity=1 ] (108.75,52) .. controls (108.75,49.65) and (110.65,47.75) .. (113,47.75) .. controls (115.35,47.75) and (117.25,49.65) .. (117.25,52) .. controls (117.25,54.35) and (115.35,56.25) .. (113,56.25) .. controls (110.65,56.25) and (108.75,54.35) .. (108.75,52) -- cycle ;
%Shape: Circle [id:dp3285031874879193] 
\draw  [fill={rgb, 255:red, 0; green, 0; blue, 0 }  ,fill opacity=1 ] (200.75,52.25) .. controls (200.75,49.9) and (202.65,48) .. (205,48) .. controls (207.35,48) and (209.25,49.9) .. (209.25,52.25) .. controls (209.25,54.6) and (207.35,56.5) .. (205,56.5) .. controls (202.65,56.5) and (200.75,54.6) .. (200.75,52.25) -- cycle ;
%Shape: Circle [id:dp9556849117327101] 
\draw  [fill={rgb, 255:red, 0; green, 0; blue, 0 }  ,fill opacity=1 ] (111,143) .. controls (111,140.65) and (112.9,138.75) .. (115.25,138.75) .. controls (117.6,138.75) and (119.5,140.65) .. (119.5,143) .. controls (119.5,145.35) and (117.6,147.25) .. (115.25,147.25) .. controls (112.9,147.25) and (111,145.35) .. (111,143) -- cycle ;
%Shape: Circle [id:dp031828947296953625] 
\draw  [fill={rgb, 255:red, 0; green, 0; blue, 0 }  ,fill opacity=1 ] (201.75,144.25) .. controls (201.75,141.9) and (203.65,140) .. (206,140) .. controls (208.35,140) and (210.25,141.9) .. (210.25,144.25) .. controls (210.25,146.6) and (208.35,148.5) .. (206,148.5) .. controls (203.65,148.5) and (201.75,146.6) .. (201.75,144.25) -- cycle ;
%Straight Lines [id:da17494894839510988] 
\draw    (113,51) -- (206,144.25) ;
%Curve Lines [id:da6183668108007425] 
\draw    (206.59,146.9) .. controls (228.24,113.54) and (252.34,159.52) .. (213.31,155.1) ;
\draw [shift={(210.84,154.76)}, rotate = 8.92] [fill={rgb, 255:red, 0; green, 0; blue, 0 }  ][line width=0.08]  [draw opacity=0] (8.93,-4.29) -- (0,0) -- (8.93,4.29) -- cycle    ;

%Straight Lines [id:da988107510408348] 
\draw [line width=3.75]    (258,95) -- (362.5,95) ;
\draw [shift={(369.5,95)}, rotate = 180] [fill={rgb, 255:red, 0; green, 0; blue, 0 }  ][line width=0.08]  [draw opacity=0] (20.54,-9.87) -- (0,0) -- (20.54,9.87) -- cycle    ;
\draw [shift={(251,95)}, rotate = 0] [fill={rgb, 255:red, 0; green, 0; blue, 0 }  ][line width=0.08]  [draw opacity=0] (20.54,-9.87) -- (0,0) -- (20.54,9.87) -- cycle    ;
%Shape: Square [id:dp1808966704275744] 
\draw  [fill={rgb, 255:red, 255; green, 255; blue, 255 }  ,fill opacity=1 ] (399,50) -- (491,50) -- (491,142) -- (399,142) -- cycle ;
%Shape: Circle [id:dp8666528918284251] 
\draw  [fill={rgb, 255:red, 0; green, 0; blue, 0 }  ,fill opacity=1 ] (394.75,51) .. controls (394.75,48.65) and (396.65,46.75) .. (399,46.75) .. controls (401.35,46.75) and (403.25,48.65) .. (403.25,51) .. controls (403.25,53.35) and (401.35,55.25) .. (399,55.25) .. controls (396.65,55.25) and (394.75,53.35) .. (394.75,51) -- cycle ;
%Shape: Circle [id:dp43877413752378036] 
\draw  [fill={rgb, 255:red, 0; green, 0; blue, 0 }  ,fill opacity=1 ] (486.75,51.25) .. controls (486.75,48.9) and (488.65,47) .. (491,47) .. controls (493.35,47) and (495.25,48.9) .. (495.25,51.25) .. controls (495.25,53.6) and (493.35,55.5) .. (491,55.5) .. controls (488.65,55.5) and (486.75,53.6) .. (486.75,51.25) -- cycle ;
%Shape: Circle [id:dp9566691290108493] 
\draw  [fill={rgb, 255:red, 0; green, 0; blue, 0 }  ,fill opacity=1 ] (397,142) .. controls (397,139.65) and (398.9,137.75) .. (401.25,137.75) .. controls (403.6,137.75) and (405.5,139.65) .. (405.5,142) .. controls (405.5,144.35) and (403.6,146.25) .. (401.25,146.25) .. controls (398.9,146.25) and (397,144.35) .. (397,142) -- cycle ;
%Shape: Circle [id:dp25039668717951924] 
\draw  [fill={rgb, 255:red, 0; green, 0; blue, 0 }  ,fill opacity=1 ] (487.75,143.25) .. controls (487.75,140.9) and (489.65,139) .. (492,139) .. controls (494.35,139) and (496.25,140.9) .. (496.25,143.25) .. controls (496.25,145.6) and (494.35,147.5) .. (492,147.5) .. controls (489.65,147.5) and (487.75,145.6) .. (487.75,143.25) -- cycle ;
%Straight Lines [id:da4840034267642028] 
\draw    (399,50) -- (492,143.25) ;
%Curve Lines [id:da6478263188731632] 
\draw    (492.59,145.9) .. controls (514.24,112.54) and (538.34,158.52) .. (499.31,154.1) ;
\draw [shift={(496.84,153.76)}, rotate = 8.92] [fill={rgb, 255:red, 0; green, 0; blue, 0 }  ][line width=0.08]  [draw opacity=0] (8.93,-4.29) -- (0,0) -- (8.93,4.29) -- cycle    ;

%Straight Lines [id:da9431928645983897] 
\draw  [dash pattern={on 0.84pt off 2.51pt}]  (6,199) -- (655-80,199) ;
%Shape: Circle [id:dp5358357890527635] 
\draw  [fill={rgb, 255:red, 0; green, 0; blue, 0 }  ,fill opacity=1 ] (103.75,237) .. controls (103.75,234.65) and (105.65,232.75) .. (108,232.75) .. controls (110.35,232.75) and (112.25,234.65) .. (112.25,237) .. controls (112.25,239.35) and (110.35,241.25) .. (108,241.25) .. controls (105.65,241.25) and (103.75,239.35) .. (103.75,237) -- cycle ;
%Shape: Circle [id:dp43707323862662983] 
\draw  [fill={rgb, 255:red, 0; green, 0; blue, 0 }  ,fill opacity=1 ] (195.75,237.25) .. controls (195.75,234.9) and (197.65,233) .. (200,233) .. controls (202.35,233) and (204.25,234.9) .. (204.25,237.25) .. controls (204.25,239.6) and (202.35,241.5) .. (200,241.5) .. controls (197.65,241.5) and (195.75,239.6) .. (195.75,237.25) -- cycle ;
%Shape: Circle [id:dp6431224706025318] 
\draw  [fill={rgb, 255:red, 0; green, 0; blue, 0 }  ,fill opacity=1 ] (106,328) .. controls (106,325.65) and (107.9,323.75) .. (110.25,323.75) .. controls (112.6,323.75) and (114.5,325.65) .. (114.5,328) .. controls (114.5,330.35) and (112.6,332.25) .. (110.25,332.25) .. controls (107.9,332.25) and (106,330.35) .. (106,328) -- cycle ;
%Shape: Circle [id:dp8511906452813223] 
\draw  [fill={rgb, 255:red, 0; green, 0; blue, 0 }  ,fill opacity=1 ] (196.75,329.25) .. controls (196.75,326.9) and (198.65,325) .. (201,325) .. controls (203.35,325) and (205.25,326.9) .. (205.25,329.25) .. controls (205.25,331.6) and (203.35,333.5) .. (201,333.5) .. controls (198.65,333.5) and (196.75,331.6) .. (196.75,329.25) -- cycle ;
%Straight Lines [id:da45378547522213697] 
\draw [line width=3.75]    (253,280) -- (357.5,280) ;
\draw [shift={(364.5,280)}, rotate = 180] [fill={rgb, 255:red, 0; green, 0; blue, 0 }  ][line width=0.08]  [draw opacity=0] (20.54,-9.87) -- (0,0) -- (20.54,9.87) -- cycle    ;
\draw [shift={(246,280)}, rotate = 0] [fill={rgb, 255:red, 0; green, 0; blue, 0 }  ][line width=0.08]  [draw opacity=0] (20.54,-9.87) -- (0,0) -- (20.54,9.87) -- cycle    ;
%Straight Lines [id:da5045261711722522] 
\draw    (108,237) -- (201,329.25) ;
%Straight Lines [id:da11638559843690044] 
\draw    (200,237.25) -- (201,329.25) ;
%Straight Lines [id:da07428374820992734] 
\draw    (108,237) -- (200,237.25) ;
%Straight Lines [id:da6143736411495485] 
\draw    (110.25,328) -- (201,329.25) ;
%Shape: Circle [id:dp17847573924704008] 
\draw  [fill={rgb, 255:red, 0; green, 0; blue, 0 }  ,fill opacity=1 ] (108,374) .. controls (108,371.65) and (109.9,369.75) .. (112.25,369.75) .. controls (114.6,369.75) and (116.5,371.65) .. (116.5,374) .. controls (116.5,376.35) and (114.6,378.25) .. (112.25,378.25) .. controls (109.9,378.25) and (108,376.35) .. (108,374) -- cycle ;
%Straight Lines [id:da3779365754281475] 
\draw    (112.25,374) -- (201,329.25) ;
%Straight Lines [id:da2958640439010536] 
\draw    (129,415) -- (201,329.25) ;
%Shape: Circle [id:dp8898700664527691] 
\draw  [fill={rgb, 255:red, 0; green, 0; blue, 0 }  ,fill opacity=1 ] (127,412) .. controls (127,409.65) and (128.9,407.75) .. (131.25,407.75) .. controls (133.6,407.75) and (135.5,409.65) .. (135.5,412) .. controls (135.5,414.35) and (133.6,416.25) .. (131.25,416.25) .. controls (128.9,416.25) and (127,414.35) .. (127,412) -- cycle ;
%Shape: Circle [id:dp14692241099198833] 
\draw  [fill={rgb, 255:red, 0; green, 0; blue, 0 }  ,fill opacity=1 ] (392.75,234) .. controls (392.75,231.65) and (394.65,229.75) .. (397,229.75) .. controls (399.35,229.75) and (401.25,231.65) .. (401.25,234) .. controls (401.25,236.35) and (399.35,238.25) .. (397,238.25) .. controls (394.65,238.25) and (392.75,236.35) .. (392.75,234) -- cycle ;
%Shape: Circle [id:dp33315416374772044] 
\draw  [fill={rgb, 255:red, 0; green, 0; blue, 0 }  ,fill opacity=1 ] (484.75,234.25) .. controls (484.75,231.9) and (486.65,230) .. (489,230) .. controls (491.35,230) and (493.25,231.9) .. (493.25,234.25) .. controls (493.25,236.6) and (491.35,238.5) .. (489,238.5) .. controls (486.65,238.5) and (484.75,236.6) .. (484.75,234.25) -- cycle ;
%Shape: Circle [id:dp6976841250455907] 
\draw  [fill={rgb, 255:red, 0; green, 0; blue, 0 }  ,fill opacity=1 ] (395,325) .. controls (395,322.65) and (396.9,320.75) .. (399.25,320.75) .. controls (401.6,320.75) and (403.5,322.65) .. (403.5,325) .. controls (403.5,327.35) and (401.6,329.25) .. (399.25,329.25) .. controls (396.9,329.25) and (395,327.35) .. (395,325) -- cycle ;
%Shape: Circle [id:dp923078230090501] 
\draw  [fill={rgb, 255:red, 0; green, 0; blue, 0 }  ,fill opacity=1 ] (485.75,326.25) .. controls (485.75,323.9) and (487.65,322) .. (490,322) .. controls (492.35,322) and (494.25,323.9) .. (494.25,326.25) .. controls (494.25,328.6) and (492.35,330.5) .. (490,330.5) .. controls (487.65,330.5) and (485.75,328.6) .. (485.75,326.25) -- cycle ;
%Straight Lines [id:da8421323532588704] 
\draw    (397,234) -- (490,326.25) ;
%Straight Lines [id:da525464383110196] 
\draw    (489,234.25) -- (490,326.25) ;
%Straight Lines [id:da39677835465487] 
\draw    (397,234) -- (489,234.25) ;
%Straight Lines [id:da5598457575196749] 
\draw    (399.25,325) -- (490,326.25) ;
%Shape: Circle [id:dp12597035613783747] 
\draw  [fill={rgb, 255:red, 0; green, 0; blue, 0 }  ,fill opacity=1 ] (397,371) .. controls (397,368.65) and (398.9,366.75) .. (401.25,366.75) .. controls (403.6,366.75) and (405.5,368.65) .. (405.5,371) .. controls (405.5,373.35) and (403.6,375.25) .. (401.25,375.25) .. controls (398.9,375.25) and (397,373.35) .. (397,371) -- cycle ;
%Straight Lines [id:da04692823881143271] 
\draw    (401.25,371) -- (490,326.25) ;
%Straight Lines [id:da34793768360898614] 
\draw    (418,412) -- (490,326.25) ;
%Shape: Circle [id:dp8724772088446058] 
\draw  [fill={rgb, 255:red, 0; green, 0; blue, 0 }  ,fill opacity=1 ] (416,409) .. controls (416,406.65) and (417.9,404.75) .. (420.25,404.75) .. controls (422.6,404.75) and (424.5,406.65) .. (424.5,409) .. controls (424.5,411.35) and (422.6,413.25) .. (420.25,413.25) .. controls (417.9,413.25) and (416,411.35) .. (416,409) -- cycle ;

% Text Node
\draw (237.55,134.78) node [anchor=north west][inner sep=0.75pt]    {$+$};
% Text Node
\draw (167,79.4) node [anchor=north west][inner sep=0.75pt]    {$-$};
% Text Node
\draw (103,30.4) node [anchor=north west][inner sep=0.75pt]    {$1$};
% Text Node
\draw (196,30.4) node [anchor=north west][inner sep=0.75pt]    {$-1$};
% Text Node
\draw (195,152.4) node [anchor=north west][inner sep=0.75pt]    {$1$};
% Text Node
\draw (98,152.4) node [anchor=north west][inner sep=0.75pt]    {$-1$};
% Text Node
\draw (386,29.4) node [anchor=north west][inner sep=0.75pt]    {$-1$};
% Text Node
\draw (485,28.4) node [anchor=north west][inner sep=0.75pt]    {$1$};
% Text Node
\draw (472,151.4) node [anchor=north west][inner sep=0.75pt]    {$-1$};
% Text Node
\draw (394,151.4) node [anchor=north west][inner sep=0.75pt]    {$1$};
% Text Node
\draw (453,78.4) node [anchor=north west][inner sep=0.75pt]    {$-$};
% Text Node
\draw (523.55,133.78) node [anchor=north west][inner sep=0.75pt]    {$+$};
% Text Node
\draw (98,215.4) node [anchor=north west][inner sep=0.75pt]    {$1$};
% Text Node
\draw (191,215.4) node [anchor=north west][inner sep=0.75pt]    {$1$};
% Text Node
\draw (199,343.4) node [anchor=north west][inner sep=0.75pt]    {$1$};
% Text Node
\draw (84,319.4) node [anchor=north west][inner sep=0.75pt]    {$1$};
% Text Node
\draw (87,368.4) node [anchor=north west][inner sep=0.75pt]    {$1$};
% Text Node
\draw (106,414.4) node [anchor=north west][inner sep=0.75pt]    {$1$};
% Text Node
\draw (387,212.4) node [anchor=north west][inner sep=0.75pt]    {$1$};
% Text Node
\draw (480,212.4) node [anchor=north west][inner sep=0.75pt]    {$1$};
% Text Node
\draw (488,340.4) node [anchor=north west][inner sep=0.75pt]    {$-1$};
% Text Node
\draw (373-10,316.4) node [anchor=north west][inner sep=0.75pt]    {$-1$};
% Text Node
\draw (373-10,365.4) node [anchor=north west][inner sep=0.75pt]    {$-1$};
% Text Node
\draw (388,411.4) node [anchor=north west][inner sep=0.75pt]    {$-1$};
% Text Node
\draw (456,372.53) node [anchor=north west][inner sep=0.75pt]    {$-$};
% Text Node
\draw (428,355.53) node [anchor=north west][inner sep=0.75pt]    {$-$};
% Text Node
\draw (416,306.53) node [anchor=north west][inner sep=0.75pt]    {$-$};
% Text Node
\draw (173,382.53) node [anchor=north west][inner sep=0.75pt]    {$-$};
% Text Node
\draw (127,337.53) node [anchor=north west][inner sep=0.75pt]    {$-$};
% Text Node
\draw (135,309.53) node [anchor=north west][inner sep=0.75pt]    {$-$};
% Text Node
\draw (146,148.4) node [anchor=north west][inner sep=0.75pt]    {$+$};
% Text Node
\draw (93,80.4) node [anchor=north west][inner sep=0.75pt]    {$+$};
% Text Node
\draw (155,30.4) node [anchor=north west][inner sep=0.75pt]    {$+$};
% Text Node
\draw (218,79.4) node [anchor=north west][inner sep=0.75pt]    {$+$};
% Text Node
\draw (438,29.4) node [anchor=north west][inner sep=0.75pt]    {$+$};
% Text Node
\draw (502,79.4) node [anchor=north west][inner sep=0.75pt]    {$+$};
% Text Node
\draw (384,79.4) node [anchor=north west][inner sep=0.75pt]    {$+$};
% Text Node
\draw (438,148.4) node [anchor=north west][inner sep=0.75pt]    {$+$};
% Text Node
\draw (145,215.4) node [anchor=north west][inner sep=0.75pt]    {$+$};
% Text Node
\draw (433,212.4) node [anchor=north west][inner sep=0.75pt]    {$+$};
% Text Node
\draw (123,266.4) node [anchor=north west][inner sep=0.75pt]    {$+$};
% Text Node
\draw (204,266.4) node [anchor=north west][inner sep=0.75pt]    {$+$};
% Text Node
\draw (407,267.4) node [anchor=north west][inner sep=0.75pt]    {$+$};
% Text Node
\draw (504,267.4) node [anchor=north west][inner sep=0.75pt]    {$+$};

\end{tikzpicture}
\caption{The asymptotic parallel (or synchronous) dynamics of two (stable) majority networks on some graphs in which the positive (resp. negative) edges are not labeled (resp are labeled by the minus symbol). In the upper panel, a majority network exhibiting a total limit cycle; in the lower panel, a majority network exhibiting a limit two-cycle. In this case, two of the nodes in the triangle are always in state $1$ and the rest of the nodes are switching.}
\label{fig:twototal}
\end{figure}

\begin{figure}[t!]
\centerline{
  \scalebox{.95}{
    %set default line width to 0.75pt        
    \tikzset{every picture/.style={line width=0.75pt}} 
    \begin{tikzpicture}[x=0.45pt,y=0.45pt,yscale=-0.75,xscale=0.75]
      %Straight Lines [id:da20043199950274515] 
      \draw (25.08,90.82) -- (114.06,90.82);
      %Straight Lines [id:da8763207823065788] 
      \draw (25.08,90.82) -- (23.8,174.18);
      %Straight Lines [id:da43915721473761926] 
      \draw (23.8,174.18) -- (112.79,174.18);
      %Straight Lines [id:da40721012592828687] 
      \draw (114.06,90.82) -- (112.79,174.18);
      %Shape: Ellipse [id:dp6902033819728984] 
      \draw [fill={rgb, 255:red, 0; green, 0; blue, 0},fill opacity=1] 
        (13,174.18) .. controls (13,168.21) and (17.84,163.36) .. 
        (23.8,163.36) .. controls (29.77,163.36) and (34.61,168.21) .. 
        (34.61,174.18) .. controls (34.61,180.16) and (29.77,185) .. 
        (23.8,185) .. controls (17.84,185) and (13,180.16) .. (13,174.18) -- 
        cycle;
      %Shape: Ellipse [id:dp3931700376781738] 
      \draw [fill={rgb, 255:red, 255; green, 255; blue, 255},fill 
        opacity=1] (101.98,174.18) .. controls (101.98,168.21) and 
        (106.82,163.36) .. (112.79,163.36) .. controls (118.75,163.36) and 
        (123.59,168.21) .. (123.59,174.18) .. controls (123.59,180.16) and 
        (118.75,185) .. (112.79,185) .. controls (106.82,185) and 
        (101.98,180.16) .. (101.98,174.18) -- cycle;
      %Shape: Ellipse [id:dp19614672596493188] 
      \draw [fill={rgb, 255:red, 255; green, 255; blue, 255},fill 
        opacity=1 ] (101.98,90.82) .. controls (101.98,84.84) and (106.82,80) 
        .. (112.79,80) .. controls (118.75,80) and (123.59,84.84) .. 
        (123.59,90.82) .. controls (123.59,96.79) and (118.75,101.64) .. 
        (112.79,101.64) .. controls (106.82,101.64) and (101.98,96.79) .. 
        (101.98,90.82) -- cycle;
      %Shape: Ellipse [id:dp5432468977566336] 
      \draw [fill={rgb, 255:red, 0; green, 0; blue, 0},fill opacity=1] 
        (14.27,90.82) .. controls (14.27,84.84) and (19.11,80) .. (25.08,80) 
        .. controls (31.04,80) and (35.88,84.84) .. (35.88,90.82) .. controls 
        (35.88,96.79) and (31.04,101.64) .. (25.08,101.64) .. controls 
        (19.11,101.64) and (14.27,96.79) .. (14.27,90.82) -- cycle;
      %Right Arrow [id:dp06779676270629953] 
      \draw (134.29,124.4) -- (163.5,124.4) -- (163.5,117.35) -- 
        (182.98,131.44) -- (163.5,145.54) -- (163.5,138.49) --  
        (134.29,138.49) -- cycle;
      %Straight Lines [id:da46324779629713664] 
      \draw (205.48,90.82) -- (294.47,90.82);
      %Straight Lines [id:da9311479787463058] 
      \draw (205.48,90.82) -- (204.21,174.18);
      %Straight Lines [id:da28823881363613446] 
      \draw (204.21,174.18) -- (293.2,174.18);
      %Straight Lines [id:da4333134417062222] 
      \draw (294.47,90.82) -- (293.2,174.18);
      %Shape: Ellipse [id:dp2581549320157651] 
      \draw [fill={rgb, 255:red, 0; green, 0; blue, 0},fill opacity=1] 
        (193.41,174.18) .. controls (193.41,168.21) and (198.25,163.36) .. 
        (204.21,163.36) .. controls (210.18,163.36) and (215.02,168.21) .. 
        (215.02,174.18) .. controls (215.02,180.16) and (210.18,185) .. 
        (204.21,185) .. controls (198.25,185) and (193.41,180.16) .. 
        (193.41,174.18) -- cycle;
      %Shape: Ellipse [id:dp9043366842819702] 
      \draw  [fill={rgb, 255:red, 255; green, 255; blue, 255},fill opacity=1] 
        (282.39,174.18) .. controls (282.39,168.21) and (287.23,163.36) .. 
        (293.2,163.36) .. controls (299.16,163.36) and (304,168.21) .. 
        (304,174.18) .. controls (304,180.16) and (299.16,185) .. (293.2,185) 
        .. controls (287.23,185) and (282.39,180.16) .. (282.39,174.18) -- 
        cycle;
      %Shape: Ellipse [id:dp2106508668776913] 
      \draw [fill={rgb, 255:red, 255; green, 255; blue, 255},fill opacity=1] 
        (282.39,90.82) .. controls (282.39,84.84) and (287.23,80) .. 
        (293.2,80) .. controls (299.16,80) and (304,84.84) .. 
        (304,90.82) .. controls (304,96.79) and (299.16,101.64) .. 
        (293.2,101.64) .. controls (287.23,101.64) and (282.39,96.79) .. 
        (282.39,90.82) -- cycle;
      %Shape: Ellipse [id:dp8139088637478189] 
      \draw  [fill={rgb, 255:red, 0; green, 0; blue, 0},fill opacity=1] 
        (194.68,90.82) .. controls (194.68,84.84) and (199.52,80) .. 
        (205.48,80) .. controls (211.45,80) and (216.29,84.84) .. 
        (216.29,90.82) .. controls (216.29,96.79) and (211.45,101.64) .. 
        (205.48,101.64) .. controls (199.52,101.64) and (194.68,96.79) .. 
        (194.68,90.82) -- cycle;

      %Straight Lines [id:da18728223148104517] 
      \draw (430.25,91.61) -- (520.5,91.61);
      %Straight Lines [id:da018429540194464522] 
      \draw (430.25,91.61) -- (428.96,173.39);
      %Straight Lines [id:da9559850041056401] 
      \draw (428.96,173.39) -- (519.21,173.39);
      %Straight Lines [id:da12954512481921487] 
      \draw (520.5,91.61) -- (519.21,173.39);
      %Shape: Ellipse [id:dp7220406997110629] 
      \draw [fill={rgb, 255:red, 0; green, 0; blue, 0},fill opacity=1] 
        (418,173.39) .. controls (418,167.53) and (422.91,162.78) .. 
        (428.96,162.78) .. controls (435.01,162.78) and (439.92,167.53) .. 
        (439.92,173.39) .. controls (439.92,179.25) and (435.01,184) .. 
        (428.96,184) .. controls (422.91,184) and (418,179.25) .. 
        (418,173.39) -- cycle;
      %Shape: Ellipse [id:dp22228475742588705] 
      \draw  [fill={rgb, 255:red, 0; green, 0; blue, 0},fill opacity=1] 
        (418,91.61) .. controls (418,85.75) and (422.91,81) .. 
        (428.96,81) .. controls (435.01,81) and (439.92,85.75) .. 
        (439.92,91.61) .. controls (439.92,97.47) and (435.01,102.22) .. 
        (428.96,102.22) .. controls (422.91,102.22) and (418,97.47) .. 
        (418,91.61) -- cycle;
      %Shape: Ellipse [id:dp883115772826589] 
      \draw [fill={rgb, 255:red, 255; green, 255; blue, 255},fill opacity=1] 
        (508,173.39) .. controls (508,167.53) and (512.91,162.78) .. 
        (518.96,162.78) .. controls (525.01,162.78) and (529.92,167.53) .. 
        (529.92,173.39) .. controls (529.92,179.25) and (525.01,184) .. 
        (518.96,184) .. controls (512.91,184) and (508,179.25) .. 
        (508,173.39) -- cycle;
      %Shape: Ellipse [id:dp026414621820427975] 
      \draw [fill={rgb, 255:red, 255; green, 255; blue, 255},fill opacity=1] 
        (508,91.61) .. controls (508,85.75) and (512.91,81) .. 
        (518.96,81) .. controls (525.01,81) and (529.92,85.75) .. 
        (529.92,91.61) .. controls (529.92,97.47) and (525.01,102.22) .. 
        (518.96,102.22) .. controls (512.91,102.22) and (508,97.47) .. 
        (508,91.61) -- cycle;
      %Right Arrow [id:dp003493420015440618] 
      \draw (540.87,124.55) -- (570.5,124.55) -- (570.5,117.64) -- 
        (590.25,131.46) -- (570.5,145.29) -- (570.5,138.38) -- 
        (540.87,138.38) -- cycle;
      %Straight Lines [id:da09878778816928147] 
      \draw (613.08,91.61) -- (703.33,91.61);
      %Straight Lines [id:da04056446285210136] 
      \draw (613.08,91.61) -- (613.08,173.39);
      %Straight Lines [id:da8742455380069465] 
      \draw (613.08,173.39) -- (703.33,173.39);
      %Straight Lines [id:da4611762495093683] 
      \draw (703.33,91.61) -- (703.33,173.39);
      %Shape: Ellipse [id:dp29662273432042896] 
      \draw [fill={rgb, 255:red, 255; green, 255; blue, 255},fill opacity=1] 
        (600.83,173.39) .. controls (600.83,167.53) and (605.74,162.78) .. 
        (611.79,162.78) .. controls (617.85,162.78) and (622.75,167.53) .. 
        (622.75,173.39) .. controls (622.75,179.25) and (617.85,184) .. 
        (611.79,184) .. controls (605.74,184) and (600.83,179.25) .. 
        (600.83,173.39) -- cycle;
      %Shape: Ellipse [id:dp33458198826666685] 
      \draw [fill={rgb, 255:red, 255; green, 255; blue, 255},fill opacity=1] 
        (600.12,91.61) .. controls (600.12,85.75) and (605.74,81) .. 
        (611.79,81) .. controls (617.85,81) and (622.75,85.75) .. 
        (622.75,91.61) .. controls (622.75,97.47) and (617.85,102.22) .. 
        (611.79,102.22) .. controls (605.74,102.22) and (600.12,97.47) .. 
        (600.12,91.61) -- cycle;
      %Shape: Ellipse [id:dp5100785720951949] 
      \draw [fill={rgb, 255:red, 0; green, 0; blue, 0},fill opacity=1] 
        (691.08,173.39) .. controls (691.08,167.53) and (695.9,162.78) .. 
        (702.04,162.78) .. controls (708.09,162.78) and (713,167.53) .. 
        (713,173.39) .. controls (713,179.25) and (708.09,184) .. 
        (702.04,184) .. controls (695.9,184) and (691.08,179.25) .. 
        (691.08,173.39) -- cycle;
      %Shape: Ellipse [id:dp45247325097315827] 
      \draw [fill={rgb, 255:red, 0; green, 0; blue, 0},fill opacity=1] 
        (691.08,91.61) .. controls (691.08,85.75) and (695.9,81) .. 
        (702.04,81) .. controls (708.09,81) and (713,85.75) .. 
        (713,91.61) .. controls (713,97.47) and (708.09,102.22) .. 
        (702.04,102.22) .. controls (695.9,102.22) and (691.08,97.47) .. 
        (691.08,91.61) -- cycle;

      %Straight Lines [id:da8392877839395512] 
      \draw [dash pattern={on 4.5pt off 4.5pt}]  (363,65) -- (363,250);

      % Text Node
      \draw (63,217) node [anchor=north west][inner sep=0.75pt]   
        [align=center] {Stable majority};
      % Text Node
      \draw (455,217) node [anchor=north west][inner sep=0.75pt]   
        [align=center] {Unstable majority};
    \end{tikzpicture}
  }
}
\caption{Example of a stable majority rule dynamics and of an 
  unstable majority rule dynamics defined on the same graph (in which all the edges are positive and not explicitly labeled): on the left panel, the dynamics reaches a fixed point; on the right panel, the dynamics reaches a limit two-cycle.}
\label{fig:stabvsunstab}
\end{figure}

\paragraph{Periodic update schemes over a threshold networks.} Let $x \in \{-1,1\}^n$ be a configuration of a threshold network $\mathcal{T}$. Let $I \subseteq V$ be a subset of vertices. The \emph{transition function} associated to $I$ is defined as the 
function:
\[
  f_I(x) = \begin{cases}
    f_i(x) & \text{if } i \in I\text{,}\\
    x_i & \text{otherwise.}
  \end{cases}
\]

Intuitively, $f_I$ assigns a new state to any node in $I$ according to its local threshold function, while the states of the nodes not in $I$ remain constant. We say that nodes in $I$ are being \emph{updated}. Given $\ell \in \N$  a \emph{periodic update scheme} is a sequence $\mu = (I_1, \ldots, I_{\ell})$ such that $I_{k} \in \mathcal{P}(V)$, where $\mathcal{P}(V)$ is the power set of $V$. The \emph{size} of a periodic update scheme is the smallest integer $r > 0$ such that $|I_k|\leq r$ for each $k \in \{1, \ldots, \ell\}$. For a configuration $x$ and an updating scheme $\mu$, we  define the \emph{global transition function} of  $\mathcal{T}$ as the function $F_\mu: \{-1,1\}^{n} \to \{-1,1\}^{n}$ given by
\[
  F_\mu(x) = \left( f_{I_\ell} \circ f_{I_{\ell-1}} \circ \ldots \circ 
    f_{I_{2}} \circ f_{I_{1}} \right) (x)\text{.}
\]

 Function $F_\mu$ assigns a new configuration for the network by sequentially applying the transition function associated to each set $I \in \mu$ according to the order defined in $\mu$. Notice that first all the nodes in $I_1$ are 
updated simultaneously, then, the ones in $I_2$ are, and so on. When the nodes in the set $I_\ell$ are updated, the next configuration of the network has been completely computed. The notions of orbit,  transient and attractors are extended to $F_\mu$ in the obvious way. To denote a threshold network $\mathcal{T} = (G,w,b)$ updated according a periodic update scheme $\mu,$ we use the notation $\mathcal{T} = (G,w,b,\mu).$ We say that the signed graph $(G,w)$ is \emph{stable for} $\mu$ if for any threshold vector $b \in \mathbb{Z}^n$ the network $\mathcal{T} = (G,w,b,\mu)$ admits only fixed points and contrarily, if there exists one threshold vector such that $\mathcal{T} = (G,w,b,\mu)$ admits limit cycles of period $s>1$ we say that $(G,w)$ is \emph{unstable} for $\mu.$
There are some important particular cases of periodic update schemes. The \emph{parallel (or synchronous) update scheme} is the scheme in which each node updates its state at the same time, i.e. $\mu = (\{V\})$. In this case, $F_\mu = F$. In the \emph{block-sequential update schemes},  $\mu$ is a partition of $V$. A \emph{sequential update schemes} is a subfamily of block-sequential update schemes is where each $I \in \mu$ is a singleton. In sequential update schemes, we can see $\mu$ as a permutation of set $V$ of the nodes of the network. 

Consider as an example the threshold network $\mathcal{T}$ defined over a cycle $C = (V, E)$, with $|V| = 4$, in which all the thresholds are fixed to $0$, i.e., $\forall i \in V, b_i = 0$, and such that the local rules correspond to the unstable majority rule. Figure~\ref{fig:update} depicts the evolution of configuration $x = (-1,1,-1,1)$ under three distinct periodic update schemes: the parallel one, the sequential one defined by $(\{1\},\{2\},\{4\},\{3\})$, and the block-sequential one defined by $(\{1,4\},\{2,3\})$. Differences of the dynamics associated to these three update schemes are emphasized. Indeed, $x$ converges to a total limit cycle under the parallel update scheme, and to a fixed point under the other update schemes.

\begin{figure}[t!]
  \centerline{
    \begin{minipage}{\textwidth}
      \centerline{\scalebox{.95}{\begin{tikzpicture}[>=latex,auto]
        \tikzstyle{node} = [circle, thick, draw]
        \node[node](n1) at (0,0) {$1$};
        \node[node](n2) at (2,0) {$2$};
        \node[node](n3) at (2,-1.5) {$3$};
        \node[node](n4) at (0,-1.5) {$4$};
        \draw[thick, -] (n1) edge [loop left] node {$-$} (n1);
        \draw[thick, -] (n2) edge [loop right] node {$-$} (n2);
        \draw[thick, -] (n3) edge [loop right] node {$-$} (n3);
        \draw[thick, -] (n4) edge [loop left] node {$-$} (n4);
        \draw[thick, -] (n1) edge node {$+$} (n2);
        \draw[thick, -] (n2) edge node {$+$} (n3);
        \draw[thick, -] (n3) edge node {$+$} (n4);
        \draw[thick, -] (n4) edge node {$+$} (n1);
      \end{tikzpicture}}}\medskip
      
      \centerline{\begin{tabular}{m{.5\textwidth}}
        \hrule
      \end{tabular}}\smallskip
    
      \centerline{\scalebox{.95}{\begin{minipage}{.300\textwidth}
        $t = 0: (-1, 1, -1, 1)$\\ %1,2,3,4
        $t = 1: (1, -1, 1, -1)$\\ %1,2,3,4
        $t = 2: (-1, 1, -1, 1)$\\ %1,2,3,4
        $t = 3: (1, -1, 1, -1)$\\ %1,2,3,4
        $t = 4: (-1, 1, -1, 1)$
      \end{minipage}
      \quad\vrule\quad
      \begin{minipage}{.300\textwidth}
        $t = 0: (-1, 1, -1, 1)$\\   %1
        $t = 1: (1, 1, -1, 1)$\\    %2
        $t = 2: (1, -1, -1, 1)$\\   %4
        $t = 3: (1, -1, -1, -1)$\\  %3
        $t = 4: (1, -1, -1, -1)$\\  %1
        $t = 5: (-1, -1, -1, -1)$
      \end{minipage}
      \quad\vrule\quad
      \begin{minipage}{.300\textwidth}
        $t = 0: (-1, 1, -1, 1)$\\   %1,4
        $t = 1: (1, 1, -1, -1)$\\   %1,3
        $t = 2: (1, -1, -1, -1)$\\   %1,4
        $t = 3: (-1, -1, -1, -1)$
      \end{minipage}}}
    \end{minipage}
  }\bigskip
  
  \centerline{
    \begin{minipage}{.2325\textwidth}
      \centerline{(a)}
    \end{minipage}
    \quad\quad
    \begin{minipage}{.274\textwidth}
      \centerline{(b)}
    \end{minipage}
    \quad\quad
    \begin{minipage}{.274\textwidth}
      \centerline{(c)}
    \end{minipage}
  }
  \caption{Threshold network based on the unstable majority rule whose 
    associated graph is depicted on the upper panel. 
    On the lower panel, three distinct evolution of configuration 
    $(-1,1,-1,1)$ according to 
    (a) the parallel update scheme, 
    (b) the sequential update scheme $(\{1\},\{2\},\{4\},\{3\})$, 
    (c) the block-sequential update scheme $(\{1,4\}, \{2,3\})$.}
  \label{fig:update}
\end{figure}

\section{Stability Index of a Graph} \label{sec:stabind} 

In this section, we introduce the stability index of a graph. To do so, we need to first introduce a concept that generalizes the notion of bipartiteness into signed graphs. 

\paragraph{Balanced and antibalanced graphs} 

The notion of a balanced signed graph was introduced by Harary~\cite{harary1953notion}. A signed graph $(G,w)$ is balanced if the vertex set of $G$ can be partitioned into two sets such that all edges with endpoints in the same partition are positive, and all edges with endpoints in different partitions are negative. More precisely, a signed graph $(G,w)$ is balanced if there exists a configuration $x\in \{-1,1\}^n$ such that, for the partition induced by the sets $V^+ = \{v \in V: x(v)=1\}$ and $V^- = \{v \in V: x(v)=-1\},$ we have that every edge $e=\{u,v\}  \in E$ satisfies:

$$w(\{u,v\}) = \begin{cases}
     1 & \text{ if } (u,v\in V^{+}) \vee (u,v \in V^{-}), \\
    -1 & \text{ otherwise. } 
\end{cases}.$$

A graph $(G,w)$ is called \emph{antibalanced} if $(G,-w)$ is balanced. 

For a signed graph $(G,w)$ we define the sign of $G$ as the product of the signs of its edges. Let $C$ be a cycle graph. We say that $C$ is an even cycle if it has an even  number of edges and otherwise we say that it is an odd cycle. 

The concept of balance and antibalance are related to the signs and the parity of the cycles of a graph \cite{zaslavsky2018negative}. In fact, a signed graph $(G,w)$ is balanced if and only if every cycle in $G$ is positive. In addition, a signed graph $(G,w)$ is antibalanced if and only if every odd cycle in $G$ is negative and every even cycle is positive. Remark that this is similar to the case of bipartite graphs in unsigned graphs which are characterized by the fact that cycles must have an even number of edges.

Since we focused in this paper on the concept of antibalance, we say that a cycle is \emph{frustrated} if it is even with negative sign or if it is odd with positive sign.

\Cref{fig:frustratedcycles} depicts different exemples of cycles which are frustrated or not.

Observe that the concepts of balance and antibalance are not complementary properties. In fact, a positive triangle is an example of a graph which is not a balanced graph nor an antibalanced graph.

\begin{figure}
\centering
\begin{tikzpicture}[x=0.55pt,y=0.55pt,yscale=-1,xscale=1]
%uncomment if require: \path (0,244); %set diagram left start at 0, and has height of 244

%Shape: Square [id:dp6086120469261367] 
\draw  [fill={rgb, 255:red, 255; green, 255; blue, 255 }  ,fill opacity=1 ] (54,64) -- (146,64) -- (146,156) -- (54,156) -- cycle ;
%Shape: Square [id:dp8297209739008786] 
\draw  [fill={rgb, 255:red, 255; green, 255; blue, 255 }  ,fill opacity=1 ] (206,65) -- (298,65) -- (298,157) -- (206,157) -- cycle ;
%Shape: Triangle [id:dp10574225672135307] 
\draw  [fill={rgb, 255:red, 255; green, 255; blue, 255 }  ,fill opacity=1 ] (404.5,65) -- (473.5,157) -- (339,157) -- cycle ;
%Shape: Triangle [id:dp3631908281172521] 
\draw  [fill={rgb, 255:red, 255; green, 255; blue, 255 }  ,fill opacity=1 ] (568.5,64) -- (637.5,156) -- (503,156) -- cycle ;
%Shape: Circle [id:dp7248412782413582] 
\draw  [fill={rgb, 255:red, 0; green, 0; blue, 0 }  ,fill opacity=1 ] (49.75,65) .. controls (49.75,62.65) and (51.65,60.75) .. (54,60.75) .. controls (56.35,60.75) and (58.25,62.65) .. (58.25,65) .. controls (58.25,67.35) and (56.35,69.25) .. (54,69.25) .. controls (51.65,69.25) and (49.75,67.35) .. (49.75,65) -- cycle ;
%Shape: Circle [id:dp33498370189581206] 
\draw  [fill={rgb, 255:red, 0; green, 0; blue, 0 }  ,fill opacity=1 ] (141.75,65.25) .. controls (141.75,62.9) and (143.65,61) .. (146,61) .. controls (148.35,61) and (150.25,62.9) .. (150.25,65.25) .. controls (150.25,67.6) and (148.35,69.5) .. (146,69.5) .. controls (143.65,69.5) and (141.75,67.6) .. (141.75,65.25) -- cycle ;
%Shape: Circle [id:dp8088136009418875] 
\draw  [fill={rgb, 255:red, 0; green, 0; blue, 0 }  ,fill opacity=1 ] (52,156) .. controls (52,153.65) and (53.9,151.75) .. (56.25,151.75) .. controls (58.6,151.75) and (60.5,153.65) .. (60.5,156) .. controls (60.5,158.35) and (58.6,160.25) .. (56.25,160.25) .. controls (53.9,160.25) and (52,158.35) .. (52,156) -- cycle ;
%Shape: Circle [id:dp8247580907854104] 
\draw  [fill={rgb, 255:red, 0; green, 0; blue, 0 }  ,fill opacity=1 ] (142.75,157.25) .. controls (142.75,154.9) and (144.65,153) .. (147,153) .. controls (149.35,153) and (151.25,154.9) .. (151.25,157.25) .. controls (151.25,159.6) and (149.35,161.5) .. (147,161.5) .. controls (144.65,161.5) and (142.75,159.6) .. (142.75,157.25) -- cycle ;
%Shape: Circle [id:dp45130855599274067] 
\draw  [fill={rgb, 255:red, 0; green, 0; blue, 0 }  ,fill opacity=1 ] (201.75,157.25) .. controls (201.75,154.9) and (203.65,153) .. (206,153) .. controls (208.35,153) and (210.25,154.9) .. (210.25,157.25) .. controls (210.25,159.6) and (208.35,161.5) .. (206,161.5) .. controls (203.65,161.5) and (201.75,159.6) .. (201.75,157.25) -- cycle ;
%Shape: Circle [id:dp6540748325685135] 
\draw  [fill={rgb, 255:red, 0; green, 0; blue, 0 }  ,fill opacity=1 ] (293.75,158.25) .. controls (293.75,155.9) and (295.65,154) .. (298,154) .. controls (300.35,154) and (302.25,155.9) .. (302.25,158.25) .. controls (302.25,160.6) and (300.35,162.5) .. (298,162.5) .. controls (295.65,162.5) and (293.75,160.6) .. (293.75,158.25) -- cycle ;
%Shape: Circle [id:dp734207280774407] 
\draw  [fill={rgb, 255:red, 0; green, 0; blue, 0 }  ,fill opacity=1 ] (201.75,65) .. controls (201.75,62.65) and (203.65,60.75) .. (206,60.75) .. controls (208.35,60.75) and (210.25,62.65) .. (210.25,65) .. controls (210.25,67.35) and (208.35,69.25) .. (206,69.25) .. controls (203.65,69.25) and (201.75,67.35) .. (201.75,65) -- cycle ;
%Shape: Circle [id:dp9966949378796987] 
\draw  [fill={rgb, 255:red, 0; green, 0; blue, 0 }  ,fill opacity=1 ] (293.75,65.25) .. controls (293.75,62.9) and (295.65,61) .. (298,61) .. controls (300.35,61) and (302.25,62.9) .. (302.25,65.25) .. controls (302.25,67.6) and (300.35,69.5) .. (298,69.5) .. controls (295.65,69.5) and (293.75,67.6) .. (293.75,65.25) -- cycle ;
%Shape: Circle [id:dp39540127056242047] 
\draw  [fill={rgb, 255:red, 0; green, 0; blue, 0 }  ,fill opacity=1 ] (400.25,66.25) .. controls (400.25,63.9) and (402.15,62) .. (404.5,62) .. controls (406.85,62) and (408.75,63.9) .. (408.75,66.25) .. controls (408.75,68.6) and (406.85,70.5) .. (404.5,70.5) .. controls (402.15,70.5) and (400.25,68.6) .. (400.25,66.25) -- cycle ;
%Shape: Circle [id:dp9517695534140526] 
\draw  [fill={rgb, 255:red, 0; green, 0; blue, 0 }  ,fill opacity=1 ] (469.25,155.75) .. controls (469.25,153.4) and (471.15,151.5) .. (473.5,151.5) .. controls (475.85,151.5) and (477.75,153.4) .. (477.75,155.75) .. controls (477.75,158.1) and (475.85,160) .. (473.5,160) .. controls (471.15,160) and (469.25,158.1) .. (469.25,155.75) -- cycle ;
%Shape: Circle [id:dp9152520180386365] 
\draw  [fill={rgb, 255:red, 0; green, 0; blue, 0 }  ,fill opacity=1 ] (334.5,157) .. controls (334.5,154.65) and (336.4,152.75) .. (338.75,152.75) .. controls (341.1,152.75) and (343,154.65) .. (343,157) .. controls (343,159.35) and (341.1,161.25) .. (338.75,161.25) .. controls (336.4,161.25) and (334.5,159.35) .. (334.5,157) -- cycle ;
%Shape: Circle [id:dp023752964100519303] 
\draw  [fill={rgb, 255:red, 0; green, 0; blue, 0 }  ,fill opacity=1 ] (564.25,64) .. controls (564.25,61.65) and (566.15,59.75) .. (568.5,59.75) .. controls (570.85,59.75) and (572.75,61.65) .. (572.75,64) .. controls (572.75,66.35) and (570.85,68.25) .. (568.5,68.25) .. controls (566.15,68.25) and (564.25,66.35) .. (564.25,64) -- cycle ;
%Shape: Circle [id:dp5101762199054278] 
\draw  [fill={rgb, 255:red, 0; green, 0; blue, 0 }  ,fill opacity=1 ] (498.75,156) .. controls (498.75,153.65) and (500.65,151.75) .. (503,151.75) .. controls (505.35,151.75) and (507.25,153.65) .. (507.25,156) .. controls (507.25,158.35) and (505.35,160.25) .. (503,160.25) .. controls (500.65,160.25) and (498.75,158.35) .. (498.75,156) -- cycle ;
%Shape: Circle [id:dp8732356724323523] 
\draw  [fill={rgb, 255:red, 0; green, 0; blue, 0 }  ,fill opacity=1 ] (633.25,156) .. controls (633.25,153.65) and (635.15,151.75) .. (637.5,151.75) .. controls (639.85,151.75) and (641.75,153.65) .. (641.75,156) .. controls (641.75,158.35) and (639.85,160.25) .. (637.5,160.25) .. controls (635.15,160.25) and (633.25,158.35) .. (633.25,156) -- cycle ;

% Text Node
\draw (30,101.4) node [anchor=north west][inner sep=0.75pt]    {$-$};
% Text Node
\draw (89,42.4) node [anchor=north west][inner sep=0.75pt]    {$+$};
% Text Node
\draw (93,163.4) node [anchor=north west][inner sep=0.75pt]    {$+$};
% Text Node
\draw (151,102.4) node [anchor=north west][inner sep=0.75pt]    {$+$};
% Text Node
\draw (182,102.4) node [anchor=north west][inner sep=0.75pt]    {$+$};
% Text Node
\draw (241,43.4) node [anchor=north west][inner sep=0.75pt]    {$+$};
% Text Node
\draw (245,164.4) node [anchor=north west][inner sep=0.75pt]    {$+$};
% Text Node
\draw (303,103.4) node [anchor=north west][inner sep=0.75pt]    {$+$};
% Text Node
\draw (346,102.4) node [anchor=north west][inner sep=0.75pt]    {$-$};
% Text Node
\draw (453,101.4) node [anchor=north west][inner sep=0.75pt]    {$+$};
% Text Node
\draw (402,163.4) node [anchor=north west][inner sep=0.75pt]    {$+$};
% Text Node
\draw (510,101.4) node [anchor=north west][inner sep=0.75pt]    {$+$};
% Text Node
\draw (617,100.4) node [anchor=north west][inner sep=0.75pt]    {$+$};
% Text Node
\draw (566,162.4) node [anchor=north west][inner sep=0.75pt]    {$+$};
% Text Node
\draw (68,187) node [anchor=north west][inner sep=0.75pt]   [align=left] {Frustrated};
% Text Node
\draw (204,187) node [anchor=north west][inner sep=0.75pt]   [align=left] {Non-frustrated};
% Text Node
\draw (355,187) node [anchor=north west][inner sep=0.75pt]   [align=left] {Non-frustrated};
% Text Node
\draw (534,186) node [anchor=north west][inner sep=0.75pt]   [align=left] {Frustrated};

\end{tikzpicture}
\caption{Some examples of frustrated and non-frustrated cycles.}
\label{fig:frustratedcycles}
\end{figure}

Let us fix a signed graph $(G,w),$ and a configuration $x\in  \{-1,1\}^n$. For each  $a,b\in \{-1,1\},$ we  define the sets of edges: $$U^{a,b}(x) = \{e=\{i,j\}\in E: x_i=x_j=a \wedge w(\{i,j\}) = b\}, \text{ and}$$ $$D^{a}(x) = \{e=\{i,j\}\in E: x_i\not = x_j \wedge w(\{i,j\}) = a\}.$$ 

Observe that the sets $$\{U^{-1,-1}(x), U^{1,-1}(x), U^{-1,1}(x), U^{1,1}(x), D^1(x), D^{-1}(x)\}$$ form a partition of the edge set of $G$ (not counting self-loops). We define the \emph{frustration index} as the quantity $$\phi(G,w) = \min\{|U^{1,-1}(x) \cup U^{-1,-1} \cup D^1(x)|, x\in \{-1,1\}^n\}.$$
Intuitively, the frustration index measures how far is a graph from being balanced. In fact, $\phi(G,w)=0$ if and only if $(G,w)$ is balanced.  This parameter was defined by Toulouse in~\cite{toulouse1987theory} as the minimum size of a set of edges $X \subseteq E$ such that $G - X$ is balanced.  Here, $G-X$ is defined as follows: if $G=(V,E)$ and $X \subseteq E$ then, $G - X = (V, E\setminus X).$  The concept of frustration has used in \cite{noual2018} to analyse the impact of asynchronism addition in the framework of Boolean networks.

From a computational standpoint, it is important to remark that computing the frustration index is $\textsc{NP}$-hard in general. In fact, it is equivalent to the ground state calculation of an Ising model without special structure and it also reduces from classic unsigned graph optimization problems such as $\textsc{EDGE BIPARTIZATION}$ and $\textsc{MAXCUT}$ \cite{aref2019balance}. 

For convenience, in this paper we work with the parameter $\rho(G, w)$, called the \emph{antifrustration index}, which corresponds to the minimum size of a set of edges $X \subseteq E$ such that $G - X$ is antibalanced. We have that $\rho(G, w) = \phi(G, -w)$. Observe that we can equivalently define \begin{equation}\label{eqn:rho}
    \rho(G,w)  =\min\{ |U^{1,1}(x)\cup U^{-1,1}(x)\cup D^{-1}(x)|, x\in \{-1,1\}^n\}.
\end{equation}

\paragraph{The stability index} We define the \emph{stability index} of a signed graph $(G,w)$ as 
the number:
\[
  \s(G,w) = -n - d^{+} + d^{-} + 2m - 4\cdot\rho(G,w)\text{.}
\]
In the next sections, we show that the stability index of a signed graph is an indicator of the possible asymptotic behaviors exhibited by the dynamics that can be defined over it.

In Figure~\ref{fig:alphavalues}, we show the value of $\s(G,w)$ for some examples of signed graphs. 

In the following we simply denote  $\rho$ and $\s$ the satisfaction index and the stability index of the corresponding signed graphs.  In the top-left panel of Figure~\ref{fig:alphavalues}, we show a signed graph with $\rho$ value is $0$ since the graph does not have any cycles. Thus, $\s = 2$. In the top-right panel, there is one frustrated cycle (one of the triangles is positive) that can be easily removed by erasing one edge, so $\rho = 1$ and $\s = 1$. In the bottom-left graph, there are two frustrated cycles; we need to remove two edges in order to have an antibalanced graph. Thus, $\rho = 2$ and $\s = -2$. Finally, in the bottom right panel, there is one frustrated cycle, this $\rho=1$ and $\s = -1$.

\begin{figure}[t!]
\centering
\tikzset{every picture/.style={line width=0.75pt}} %set default line width to 0.75pt        

\begin{tikzpicture}[x=0.5pt,y=0.5pt,yscale=-1,xscale=1]
%uncomment if require: \path (0,419); %set diagram left start at 0, and has height of 419

%Shape: Triangle [id:dp6611936813954167] 
\draw  [fill={rgb, 255:red, 255; green, 255; blue, 255 }  ,fill opacity=1 ] (157.84,276.91) -- (245.58,218.47) -- (245.35,332.74) -- cycle ;
%Shape: Triangle [id:dp8864114643112826] 
\draw  [fill={rgb, 255:red, 255; green, 255; blue, 255 }  ,fill opacity=1 ] (471.97,262.47) -- (530.59,350.09) -- (416.32,350.09) -- cycle ;
%Shape: Ellipse [id:dp7490985180915135] 
\draw  [fill={rgb, 255:red, 0; green, 0; blue, 0 }  ,fill opacity=1 ] (526.98,348.9) .. controls (526.98,346.67) and (528.6,344.86) .. (530.59,344.86) .. controls (532.59,344.86) and (534.21,346.67) .. (534.21,348.9) .. controls (534.21,351.14) and (532.59,352.95) .. (530.59,352.95) .. controls (528.6,352.95) and (526.98,351.14) .. (526.98,348.9) -- cycle ;
%Shape: Ellipse [id:dp12669461837520735] 
\draw  [fill={rgb, 255:red, 0; green, 0; blue, 0 }  ,fill opacity=1 ] (412.5,350.09) .. controls (412.5,347.86) and (414.12,346.05) .. (416.11,346.05) .. controls (418.11,346.05) and (419.72,347.86) .. (419.72,350.09) .. controls (419.72,352.33) and (418.11,354.14) .. (416.11,354.14) .. controls (414.12,354.14) and (412.5,352.33) .. (412.5,350.09) -- cycle ;
%Shape: Ellipse [id:dp10995179580350722] 
\draw  [fill={rgb, 255:red, 0; green, 0; blue, 0 }  ,fill opacity=1 ] (468.36,262.71) .. controls (468.36,260.47) and (469.98,258.66) .. (471.97,258.66) .. controls (473.97,258.66) and (475.58,260.47) .. (475.58,262.71) .. controls (475.58,264.94) and (473.97,266.76) .. (471.97,266.76) .. controls (469.98,266.76) and (468.36,264.94) .. (468.36,262.71) -- cycle ;
%Curve Lines [id:da5038587733692923] 
\draw    (471.97,258.66) .. controls (460.32,218.53) and (500.06,226.87) .. (481.67,261.28) ;
\draw [shift={(480.47,263.42)}, rotate = 300.41] [fill={rgb, 255:red, 0; green, 0; blue, 0 }  ][line width=0.08]  [draw opacity=0] (8.93,-4.29) -- (0,0) -- (8.93,4.29) -- cycle    ;
%Curve Lines [id:da48814640947655574] 
\draw    (530.59,348.9) .. controls (552.24,315.54) and (576.34,361.52) .. (537.31,357.1) ;
\draw [shift={(534.84,356.76)}, rotate = 8.92] [fill={rgb, 255:red, 0; green, 0; blue, 0 }  ][line width=0.08]  [draw opacity=0] (8.93,-4.29) -- (0,0) -- (8.93,4.29) -- cycle    ;

%Shape: Square [id:dp10405657319718631] 
\draw  [fill={rgb, 255:red, 255; green, 255; blue, 255 }  ,fill opacity=1 ] (420,29) -- (512,29) -- (512,121) -- (420,121) -- cycle ;
%Shape: Circle [id:dp22190377957238172] 
\draw  [fill={rgb, 255:red, 0; green, 0; blue, 0 }  ,fill opacity=1 ] (415.75,30) .. controls (415.75,27.65) and (417.65,25.75) .. (420,25.75) .. controls (422.35,25.75) and (424.25,27.65) .. (424.25,30) .. controls (424.25,32.35) and (422.35,34.25) .. (420,34.25) .. controls (417.65,34.25) and (415.75,32.35) .. (415.75,30) -- cycle ;
%Shape: Circle [id:dp10678580088208944] 
\draw  [fill={rgb, 255:red, 0; green, 0; blue, 0 }  ,fill opacity=1 ] (507.75,30.25) .. controls (507.75,27.9) and (509.65,26) .. (512,26) .. controls (514.35,26) and (516.25,27.9) .. (516.25,30.25) .. controls (516.25,32.6) and (514.35,34.5) .. (512,34.5) .. controls (509.65,34.5) and (507.75,32.6) .. (507.75,30.25) -- cycle ;
%Shape: Circle [id:dp63615997529201] 
\draw  [fill={rgb, 255:red, 0; green, 0; blue, 0 }  ,fill opacity=1 ] (418,121) .. controls (418,118.65) and (419.9,116.75) .. (422.25,116.75) .. controls (424.6,116.75) and (426.5,118.65) .. (426.5,121) .. controls (426.5,123.35) and (424.6,125.25) .. (422.25,125.25) .. controls (419.9,125.25) and (418,123.35) .. (418,121) -- cycle ;
%Shape: Circle [id:dp04507112844555006] 
\draw  [fill={rgb, 255:red, 0; green, 0; blue, 0 }  ,fill opacity=1 ] (508.75,122.25) .. controls (508.75,119.9) and (510.65,118) .. (513,118) .. controls (515.35,118) and (517.25,119.9) .. (517.25,122.25) .. controls (517.25,124.6) and (515.35,126.5) .. (513,126.5) .. controls (510.65,126.5) and (508.75,124.6) .. (508.75,122.25) -- cycle ;
%Straight Lines [id:da4008097424112578] 
\draw    (420,29) -- (513,122.25) ;
%Curve Lines [id:da08026140631144996] 
\draw    (513.59,124.9) .. controls (535.24,91.54) and (559.34,137.52) .. (520.31,133.1) ;
\draw [shift={(517.84,132.76)}, rotate = 8.92] [fill={rgb, 255:red, 0; green, 0; blue, 0 }  ][line width=0.08]  [draw opacity=0] (8.93,-4.29) -- (0,0) -- (8.93,4.29) -- cycle    ;

%Straight Lines [id:da9310325716820926] 
\draw    (315,2) -- (315.5,423) ;
%Straight Lines [id:da2565669484850799] 
\draw    (0,201) -- (661.5,201) ;
%Shape: Triangle [id:dp022286489226140294] 
\draw  [fill={rgb, 255:red, 255; green, 255; blue, 255 }  ,fill opacity=1 ] (158.26,277.37) -- (71.15,336.76) -- (70.15,222.49) -- cycle ;
%Shape: Ellipse [id:dp9198086109354282] 
\draw  [fill={rgb, 255:red, 0; green, 0; blue, 0 }  ,fill opacity=1 ] (72.31,333.14) .. controls (74.55,333.12) and (76.37,334.72) .. (76.39,336.71) .. controls (76.41,338.71) and (74.61,340.34) .. (72.37,340.36) .. controls (70.14,340.38) and (68.31,338.78) .. (68.29,336.78) .. controls (68.28,334.79) and (70.08,333.16) .. (72.31,333.14) -- cycle ;
%Shape: Ellipse [id:dp5847753861446787] 
\draw  [fill={rgb, 255:red, 0; green, 0; blue, 0 }  ,fill opacity=1 ] (70.12,218.67) .. controls (72.36,218.65) and (74.18,220.25) .. (74.2,222.24) .. controls (74.22,224.24) and (72.42,225.87) .. (70.18,225.89) .. controls (67.95,225.91) and (66.12,224.31) .. (66.1,222.32) .. controls (66.09,220.32) and (67.89,218.69) .. (70.12,218.67) -- cycle ;
%Shape: Ellipse [id:dp2150591955385479] 
\draw  [fill={rgb, 255:red, 0; green, 0; blue, 0 }  ,fill opacity=1 ] (157.99,273.77) .. controls (160.23,273.75) and (162.05,275.35) .. (162.07,277.34) .. controls (162.09,279.34) and (160.29,280.97) .. (158.05,280.99) .. controls (155.82,281.01) and (153.99,279.41) .. (153.97,277.41) .. controls (153.96,275.42) and (155.75,273.79) .. (157.99,273.77) -- cycle ;
%Curve Lines [id:da08292542639207778] 
\draw    (72.34,336.75) .. controls (105.9,358.11) and (60.13,382.6) .. (64.21,343.54) ;
\draw [shift={(64.52,341.07)}, rotate = 98.42] [fill={rgb, 255:red, 0; green, 0; blue, 0 }  ][line width=0.08]  [draw opacity=0] (8.93,-4.29) -- (0,0) -- (8.93,4.29) -- cycle    ;
%Shape: Ellipse [id:dp8675098724989143] 
\draw  [fill={rgb, 255:red, 0; green, 0; blue, 0 }  ,fill opacity=1 ] (245.39,222.07) .. controls (243.15,222.07) and (241.34,220.45) .. (241.35,218.46) .. controls (241.35,216.46) and (243.17,214.85) .. (245.4,214.85) .. controls (247.64,214.86) and (249.45,216.48) .. (249.44,218.47) .. controls (249.44,220.47) and (247.62,222.08) .. (245.39,222.07) -- cycle ;
%Shape: Ellipse [id:dp8149671144625001] 
\draw  [fill={rgb, 255:red, 0; green, 0; blue, 0 }  ,fill opacity=1 ] (245.34,336.56) .. controls (243.1,336.56) and (241.29,334.94) .. (241.3,332.94) .. controls (241.3,330.95) and (243.12,329.33) .. (245.35,329.34) .. controls (247.59,329.34) and (249.4,330.96) .. (249.39,332.96) .. controls (249.39,334.95) and (247.57,336.57) .. (245.34,336.56) -- cycle ;
%Straight Lines [id:da897937439299864] 
\draw    (112,93) -- (180.5,93) ;
\draw [shift={(180.5,93)}, rotate = 0] [color={rgb, 255:red, 0; green, 0; blue, 0 }  ][fill={rgb, 255:red, 0; green, 0; blue, 0 }  ][line width=0.75]      (0, 0) circle [x radius= 3.35, y radius= 3.35]   ;
\draw [shift={(112,93)}, rotate = 0] [color={rgb, 255:red, 0; green, 0; blue, 0 }  ][fill={rgb, 255:red, 0; green, 0; blue, 0 }  ][line width=0.75]      (0, 0) circle [x radius= 3.35, y radius= 3.35]   ;
%Straight Lines [id:da2714686785499312] 
\draw    (180.5,93) -- (239.5,123) ;
\draw [shift={(239.5,123)}, rotate = 26.95] [color={rgb, 255:red, 0; green, 0; blue, 0 }  ][fill={rgb, 255:red, 0; green, 0; blue, 0 }  ][line width=0.75]      (0, 0) circle [x radius= 3.35, y radius= 3.35]   ;

% Text Node
\draw (546.55,115.78) node [anchor=north west][inner sep=0.75pt]    {$+$};
% Text Node
\draw (474,57.4) node [anchor=north west][inner sep=0.75pt]    {$-$};
% Text Node
\draw (370,162.4) node [anchor=north west][inner sep=0.75pt]  [font=\small]  {$\s =-4-1+0+2\times 5-4\times 1=1\ $};
% Text Node
\draw (527.55,65.78) node [anchor=north west][inner sep=0.75pt]    {$+$};
% Text Node
\draw (458.55,4.78) node [anchor=north west][inner sep=0.75pt]    {$-$};
% Text Node
\draw (388.55,66.78) node [anchor=north west][inner sep=0.75pt]    {$+$};
% Text Node
\draw (459.55,130.78) node [anchor=north west][inner sep=0.75pt]    {$+$};
% Text Node
\draw (50-55,387.02) node [anchor=north west][inner sep=0.75pt]  [font=\small]  {$\s =-5-1+0+2\times 6-4\times 2=-2\ $};
% Text Node
\draw (50-40,160.02) node [anchor=north west][inner sep=0.75pt]  [font=\small]  {$\s =-2-0+0+2\times 2-4\times 0=2\ $};
% Text Node
\draw (428.09,291.06) node [anchor=north west][inner sep=0.75pt]    {$+$};
% Text Node
\draw (508.8,291.06) node [anchor=north west][inner sep=0.75pt]    {$+$};
% Text Node
\draw (468.87,354.88) node [anchor=north west][inner sep=0.75pt]    {$+$};
% Text Node
\draw (492.66,235.82) node [anchor=north west][inner sep=0.75pt]    {$+$};
% Text Node
\draw (561.55,336.78) node [anchor=north west][inner sep=0.75pt]    {$-$};
% Text Node
\draw (57.83,268.08) node [anchor=north west][inner sep=0.75pt]  [rotate=-89.5]  {$+$};
% Text Node
\draw (103.99,359.45) node [anchor=north west][inner sep=0.75pt]  [rotate=-89.5]  {$+$};
% Text Node
\draw (186.34,320.85) node [anchor=north west][inner sep=0.75pt]  [rotate=-270.12]  {$+$};
% Text Node
\draw (186.51,240.14) node [anchor=north west][inner sep=0.75pt]  [rotate=-270.12]  {$+$};
% Text Node
\draw (250.24,280.2) node [anchor=north west][inner sep=0.75pt]  [rotate=-270.12]  {$+$};
% Text Node
\draw (116.34,320.85) node [anchor=north west][inner sep=0.75pt]  [rotate=-270.12]  {$+$};
% Text Node
\draw (116.34,239.85) node [anchor=north west][inner sep=0.75pt]  [rotate=-270.12]  {$+$};
% Text Node
\draw (371,386.4) node [anchor=north west][inner sep=0.75pt]  [font=\small]  {$\s =-3-1+1+2\times 3-4\times 1=-1\ $};
% Text Node
\draw (130.2,66.07) node [anchor=north west][inner sep=0.75pt]  [rotate=-1.13]  {$+$};
% Text Node
\draw (218.3,84.32) node [anchor=north west][inner sep=0.75pt]  [rotate=-28.46]  {$-$};
\end{tikzpicture}

\caption{Values for the stability index in different graphs. top-left:  $\rho = 0$; top-right $\rho = 1$; bottom-left $\rho = 2$; bottom-right $\rho=1$.}
\label{fig:alphavalues}
\end{figure}

\subsection{Existence of total limit cycles imply positive stability index }

We now show the following relation between the stability index and the existence of total limit cycles. 

\begin{theorem}\label{theo:totalcycles}
    Let $\mathcal{T} = (G,w,b)$ be a threshold network that admits a total limit cycle. Then $\s(G,w) \geq 0$. 
\end{theorem}

Before proving \Cref{theo:totalcycles} we give some technical lemmas. The first one states that when a threshold network defined over a signed graph $(G,w)$ admits total limit cycle, then the associated majority networks, i.e. the threshold network defined over $(G,w)$ whose threshold vector is $\vv{0} = (0, \dots, 0)$, admits a total limit cycle as well

\begin{lemma}\label{lem:totalcycles}
    Let $\mathcal{T} = (G,w,b)$ be a threshold network that admits a total limit cycle. Then $\widehat{\mathcal{T}} = (G,w,\vv{0})$  admits a total limit cycle as well.
\end{lemma}

\begin{proof}
    Let $x$ be a configuration defining a total limit cycle. Then $x'=F(x) = -x$. Let $i$ be an arbitrary node. Without loss of generality we assume that $x_i = 1$. Since $f_i(x) = -1$ we have that

    $$\sum_{j\in V} w_{ij} x_j - b_i <0 \Rightarrow \sum_{j\in V} w_{ij} x_j  <b_i,$$
    and since $f_i(x') = 1$
    $$\sum_{j\in V} w_{ij} x'_j - b_i >0 \Rightarrow -\sum_{j\in V} w_{ij} x_j - b_i >0 \Rightarrow \sum_{j\in V} w_{ij} x_j < -b_i, $$
    We deduce that $$\sum_{j\in V} w_{ij} x_j < -|b_i| \leq 0 $$  and $$\sum_{j\in V} w_{ij} x'_j >  0. $$ 
    Then, we can switch the threshold $b_i$ of $i$ to $0$ and the node still oscillates. 
\end{proof}

Next lemma provides a characterization of the configurations that reach a total limit cycle.

\begin{lemma} 
  \label{lemma:psws}
The threshold network $\mathcal{T} = (G,w,\vv{0})$   admits a total limit cycle if and only if there exists a 
  configuration $x \in \{-1,1\}^{n}$ such that, for all $i \in V$ 
$$\left(  \sum_{j\neq i} w_{ij}x_ix_j +  w_{ii} + 1 \right) \leq 0.$$  
\end{lemma}

\begin{proof}
Suppose first that $\mathcal{T}$ admits a total limit cycle. We pick as $x$ any configuration such that $x'=F(x) = -x$.   Suppose that $x_i=1$. Then $x'_i = -1$, which implies
\begin{align*}\sum_{j\in V} w_{ij}x_j  < 0
\iff & \sum_{j\neq i} w_{ij}x_j + w_{ii}x_i < 0 \\
\iff & \sum_{j\neq i} w_{ij}x_ix_j + w_{ii}  < 0 \\
\iff & \sum_{j\neq i} w_{ij}x_ix_j + w_{ii} + 1\leq 0. 
\end{align*}
Now suppose that $x_i = -1$. Then $x'_i = 1$, which implies 
\begin{align*}\sum_{j\in V} w_{ij}x_j  > 0
\iff & \sum_{j\neq i} w_{ij}x_j + w_{ii}x_i > 0 \\
\iff & -\sum_{j\neq i} w_{ij}x_ix_j - w_{ii} > 0 \\
\iff & \sum_{j\neq i} w_{ij}x_ix_j + w_{ii} < 0 \\
\iff & \sum_{j\neq i} w_{ij}x_ix_j + w_{ii} +1 \leq  0. 
\end{align*}
The reciprocal follows  from the same calculations. 
\end{proof}

The next lemma provides an alternative representation of the stability index. Let $(G,w)$ be an arbitrary signed graph and let $y\in \{-1,1\}^n$ be a configuration of $(G,w)$. We define the parameter $$\varphi(y) = \dfrac{1}{2}\sum_{i \in [n]}\sum_{j\neq i}  w_{ij}y_iy_j.$$

\begin{lemma}\label{lem:varphi}
Let $(G,w)$ be an arbitrary signed graph. Then $$\min_{y\in \{-1,1\}^n}\varphi(y) =  2\rho(G,w)-m.$$
Furthermore, 
$$\s(G,w) = -n -d^+ + d^- - 2\left(\min_{x\in\{-1,1\}^n}\varphi(x)\right).$$
\end{lemma}

\begin{proof}
    First, notice that
\begin{align*}
\dfrac{1}{2}\sum_{i \in [n]}\sum_{j\neq i}  w_{ij}y_iy_j = \sum_{\{i,j\}\in D^{+1}(y)}w_{ij} y_iy_j &+ \sum_{\{i,j\}\in D^{-1}(y)}w_{ij} y_iy_j\\ 
&+ \sum_{a,b \in \{-1,1\}}\sum_{\{i,j\}\in U^{a,b}(y)}w_{ij} y_iy_j.
\end{align*}
and also
\begin{align*}
    \sum_{\{i,j\}\in U^{1,1}(y)}w_{ij} y_iy_j =& |U^{1,1}(y)|\\
    \sum_{\{i,j\}\in U^{-1,1}(y)}w_{ij} y_iy_j =& |U^{-1,1}(y)|\\
    \sum_{\{i,j\}\in U^{1,-1}(y)}w_{ij} y_iy_j =& -|U^{1,-1}(y)|\\
    \sum_{\{i,j\}\in U^{-1,-1}(y)}w_{ij} y_iy_j =& -|U^{-1,-1}(y)|\\
    \sum_{\{i,j\}\in D^{1}(y)}w_{ij} y_iy_j =& -|D^{1}(y)|\\
    \sum_{\{i,j\}\in D^{-1}(y)}w_{ij} y_iy_j =& |D^{-1}(y)|.
\end{align*}
Hence, 
\begin{align*}
\varphi(y) 
= (|U^{1,1}(y)| + |U^{-1,1}(y)| & + |D^{-1}(y)|)\\& - (|U^{1,-1}(y)| + |U^{-1,-1}(y)| + |D^{1}(y)|).
\end{align*}
Now using that
$$m = |U^{1,1}(y)| + |U^{-1,1}(y)| + |D^{-1}(y)| + |U^{1,-1}(y)| + |U^{-1,-1}(y)| + |D^{1}(y)|,$$
we obtain
$$\varphi(y) = 2(|U^{+1,+1}(y)| + |U^{-1,+1}(y)| + |D^{-1}(y)|) - m.$$
Finally, by using \Cref{eqn:rho} we deduce that $$\min_{y\in \{-1,1\}^n} \varphi(y) =  2\rho(G,w) - m.$$

     Now remember that $$\s(G,w) = -n - d^+ + d^- +2m - 4\rho(G,w).$$ Since $$2m - 4\rho(G,w) = -2\left(\min_{x\in\{-1,1\}^n}\varphi(x)\right), $$ we deduce that $$\s(G,w) =  -n -d^+ + d^- - 2\left(\min_{x\in\{-1,1\}^n}\varphi(x)\right).$$
\end{proof}

We are now ready to show \Cref{theo:totalcycles}.

\begin{proof}[Proof of \Cref{theo:totalcycles}]
\Cref{lem:totalcycles} implies that we can assume that $\mathcal{T} = (G,w,0)$.  Let $x$ be such that $F(x) = -x$. From \Cref{lemma:psws} we have that for every $i\in V$, $$\left(  \sum_{j\neq i} w_{ij}x_ix_j +  w_{ii} + 1 \right) \leq 0.$$  
Then,
$$
\sum_{i\in [n]}\left(  \sum_{j\neq i} w_{ij}x_ix_j +  w_{ii} + 1 \right) \leq 0  \Rightarrow 2\varphi(x) + d^+ - d^- + n \leq 0. $$

Finally, using \Cref{lem:varphi} we deduce that $  \s(G,w) \geq 0.$
\end{proof}

\begin{remark}
    Observe that  studying only the sign of the stability index $\s(G,w)$ is not sufficient to deduce if a network exhibits a total limit cycle. In fact, as we show in Figure \ref{fig:ejemplociclos} there are examples of graphs having  $\s(G,w) \geq 0$ which do not admit total limit cycles.
\end{remark}

\begin{figure}
    \centering

\tikzset{every picture/.style={line width=0.75pt}} %set default line width to 0.75pt        

\begin{tikzpicture}[x=0.66pt,y=0.55pt,yscale=-1,xscale=1]

%Shape: Triangle [id:dp8980244092728822] 
\draw  [fill={rgb, 255:red, 255; green, 255; blue, 255 }  ,fill opacity=1 ] (191.5,59) -- (260.5,151) -- (126,151) -- cycle ;
%Shape: Circle [id:dp444036920245207] 
\draw  [fill={rgb, 255:red, 0; green, 0; blue, 0 }  ,fill opacity=1 ] (187.25,59) .. controls (187.25,56.65) and (189.15,54.75) .. (191.5,54.75) .. controls (193.85,54.75) and (195.75,56.65) .. (195.75,59) .. controls (195.75,61.35) and (193.85,63.25) .. (191.5,63.25) .. controls (189.15,63.25) and (187.25,61.35) .. (187.25,59) -- cycle ;
%Shape: Circle [id:dp6567713303573612] 
\draw  [fill={rgb, 255:red, 0; green, 0; blue, 0 }  ,fill opacity=1 ] (121.75,151) .. controls (121.75,148.65) and (123.65,146.75) .. (126,146.75) .. controls (128.35,146.75) and (130.25,148.65) .. (130.25,151) .. controls (130.25,153.35) and (128.35,155.25) .. (126,155.25) .. controls (123.65,155.25) and (121.75,153.35) .. (121.75,151) -- cycle ;
%Shape: Circle [id:dp5025867638899473] 
\draw  [fill={rgb, 255:red, 0; green, 0; blue, 0 }  ,fill opacity=1 ] (256.25,151) .. controls (256.25,148.65) and (258.15,146.75) .. (260.5,146.75) .. controls (262.85,146.75) and (264.75,148.65) .. (264.75,151) .. controls (264.75,153.35) and (262.85,155.25) .. (260.5,155.25) .. controls (258.15,155.25) and (256.25,153.35) .. (256.25,151) -- cycle ;
%Curve Lines [id:da45424930600363256] 
\draw    (189.97,51.66) .. controls (178.32,11.53) and (218.06,19.87) .. (199.67,54.28) ;
\draw [shift={(198.47,56.42)}, rotate = 300.41] [fill={rgb, 255:red, 0; green, 0; blue, 0 }  ][line width=0.08]  [draw opacity=0] (8.93,-4.29) -- (0,0) -- (8.93,4.29) -- cycle    ;
%Straight Lines [id:da3298756969140906] 
\draw    (353.5,0) -- (356.5,298) ;
%Straight Lines [id:da9335347178653655] 
\draw    (471,195) -- (518.5,195) ;
\draw [shift={(521.5,195)}, rotate = 180] [fill={rgb, 255:red, 0; green, 0; blue, 0 }  ][line width=0.08]  [draw opacity=0] (8.93,-4.29) -- (0,0) -- (8.93,4.29) -- cycle    ;
\draw [shift={(468,195)}, rotate = 0] [fill={rgb, 255:red, 0; green, 0; blue, 0 }  ][line width=0.08]  [draw opacity=0] (8.93,-4.29) -- (0,0) -- (8.93,4.29) -- cycle    ;
%Straight Lines [id:da3737852157869057] 
\draw    (464,145) -- (511.5,145) ;
\draw [shift={(514.5,145)}, rotate = 180] [fill={rgb, 255:red, 0; green, 0; blue, 0 }  ][line width=0.08]  [draw opacity=0] (8.93,-4.29) -- (0,0) -- (8.93,4.29) -- cycle    ;
\draw [shift={(461,145)}, rotate = 0] [fill={rgb, 255:red, 0; green, 0; blue, 0 }  ][line width=0.08]  [draw opacity=0] (8.93,-4.29) -- (0,0) -- (8.93,4.29) -- cycle    ;
%Straight Lines [id:da10997548855873895] 
\draw    (461,103) -- (508.5,103) ;
\draw [shift={(511.5,103)}, rotate = 180] [fill={rgb, 255:red, 0; green, 0; blue, 0 }  ][line width=0.08]  [draw opacity=0] (8.93,-4.29) -- (0,0) -- (8.93,4.29) -- cycle    ;
\draw [shift={(458,103)}, rotate = 0] [fill={rgb, 255:red, 0; green, 0; blue, 0 }  ][line width=0.08]  [draw opacity=0] (8.93,-4.29) -- (0,0) -- (8.93,4.29) -- cycle    ;
%Straight Lines [id:da393715151271687] 
\draw    (460,64) -- (507.5,64) ;
\draw [shift={(510.5,64)}, rotate = 180] [fill={rgb, 255:red, 0; green, 0; blue, 0 }  ][line width=0.08]  [draw opacity=0] (8.93,-4.29) -- (0,0) -- (8.93,4.29) -- cycle    ;
\draw [shift={(457,64)}, rotate = 0] [fill={rgb, 255:red, 0; green, 0; blue, 0 }  ][line width=0.08]  [draw opacity=0] (8.93,-4.29) -- (0,0) -- (8.93,4.29) -- cycle    ;

% Text Node
\draw    (370,54) -- (441+5,54) -- (441+5,76) -- (370,76) -- cycle  ;
\draw (373,58.4) node [anchor=north west][inner sep=0.75pt]  [font=\small]  {$-1-1-1$};
% Text Node
\draw    (388,93) -- (444,93) -- (444,115) -- (388,115) -- cycle  ;
\draw (391,97.4) node [anchor=north west][inner sep=0.75pt]  [font=\small]  {$-1-11$};
% Text Node
\draw    (522,53) -- (578,53) -- (578,75) -- (522,75) -- cycle  ;
\draw (525,57.4) node [anchor=north west][inner sep=0.75pt]  [font=\small]  {$-11-1$};
% Text Node
\draw    (520,92) -- (560,92) -- (560,114) -- (520,114) -- cycle  ;
\draw (523,96.4) node [anchor=north west][inner sep=0.75pt]  [font=\small]  {$-111$};
% Text Node
\draw    (388,133) -- (450,133) -- (450,155) -- (388,155) -- cycle  ;
\draw (391,137.4) node [anchor=north west][inner sep=0.75pt]  [font=\small]  {$1-1-1$};
% Text Node
\draw    (412,180) -- (458,180) -- (458,202) -- (412,202) -- cycle  ;
\draw (415,184.4) node [anchor=north west][inner sep=0.75pt]  [font=\small]  {$1-11$};
% Text Node
\draw    (520,133) -- (566,133) -- (566,155) -- (520,155) -- cycle  ;
\draw (523,137.4) node [anchor=north west][inner sep=0.75pt]  [font=\small]  {$11-1$};
% Text Node
\draw    (526,180) -- (557,180) -- (557,202) -- (526,202) -- cycle  ;
\draw (529,184.4) node [anchor=north west][inner sep=0.75pt]  [font=\small]  {$111$};
% Text Node
\draw    (56,174) -- (130,174) -- (130,198) -- (56,198) -- cycle  ;
\draw (59,178.4) node [anchor=north west][inner sep=0.75pt]  [font=\small]  {$b_{1} \ =\ -2$};
% Text Node
\draw    (248,35) -- (307,35) -- (307,59) -- (248,59) -- cycle  ;
\draw (251,39.4) node [anchor=north west][inner sep=0.75pt]  [font=\small]  {$b_{2} \ =\ 2$};
% Text Node
\draw    (263,175) -- (337,175) -- (337,199) -- (263,199) -- cycle  ;
\draw (266,179.4) node [anchor=north west][inner sep=0.75pt]  [font=\small]  {$b _{3} \ =\ -2$};
% Text Node
\draw (133,96.4) node [anchor=north west][inner sep=0.75pt]    {$+$};
% Text Node
\draw (240,95.4) node [anchor=north west][inner sep=0.75pt]    {$+$};
% Text Node
\draw (189,157.4) node [anchor=north west][inner sep=0.75pt]    {$+$};
% Text Node
\draw (210.66,28.82) node [anchor=north west][inner sep=0.75pt]    {$-$};
% Text Node
\draw (34,237.4) node [anchor=north west][inner sep=0.75pt]  [font=\small]  {$\mathcal{S}( G) \ =\ -3\ -0\ +1\ +2\times 3\ -\ 4\times 1\ =\ 0\ \geq 0\ $};
% Text Node
\draw (100,146.4) node [anchor=north west][inner sep=0.75pt]  [font=\small]  {$1$};
% Text Node
\draw (163,48.4) node [anchor=north west][inner sep=0.75pt]  [font=\small]  {$2$};
% Text Node
\draw (276,148.4) node [anchor=north west][inner sep=0.75pt]  [font=\small]  {$3$};

\end{tikzpicture}
    \caption{(Left Panel) An example of a network with non-negative stability index which does not admit total two-cycles (updated in parallel). (Right Panel) Dynamics of the example in the right panel. }
    \label{fig:ejemplociclos}
\end{figure}

\section{Stability of Synchronous Dynamics}
\label{sec:par}

In this section, we study the link between the stability index of a signed graph $(G,w)$ and the synchronous dynamics induced by any threshold network defined over it. In particular, we give necessarily and sufficient conditions for stability. We focus in this section on the synchronous dynamics, i.e., the ones induced by parallel update schemes, and we focus on the existence of fixed points as only attractors of the dynamics and continue with the existence of total two cycles. 

\subsection{A sufficient condition for stability}
We begin by stating \Cref{thm:goles1,thm:goles2}, which are well known results on  threshold networks. 

\begin{proposition}[\cite{goles1980}]
  \label{thm:goles1}
  Let  $\mathcal{T} =(G,w,b)$ threshold network of size $n$. For every $x \in \{-1,1\}^{n}$, there exists $t \in \mathbb{N}$ such   that $F^t(x) = F^{t+2}(x)$.
\end{proposition}
In full words, \Cref{thm:goles1} states that every initial configuration of a threshold network converges to either a fixed point or a limit two-cycle. 

Next proposition gives a sufficient condition for having only fixed points (i.e. only limit cycles of period $1$). We say that a  matrix $M$ of order $n$ is \emph{positive semi-definite} on $\{0,1\}^n$ if $x^T W x \geq 0$ for all $x \in \{-1,1\}^{n}$.

\begin{proposition}[\cite{goles1982}]
  \label{thm:goles2}
  Let $\mathcal{T} = (G,w,b)$ be a threshold 
  network of size $n$ such that $W=(w_{ij})$ is positive semi-definite. Then $\mathcal{T}$ is stable. 
\end{proposition}

The proof of \Cref{thm:goles1,thm:goles2} use the definition of the following \emph{energy functional}:
\[
  L(x) = - \frac{1}{2} x^\mathrm{T} W x + b^\mathrm{T} x\text{.}
\]
More precisely, it uses the following property of the energy functional. 
\begin{proposition}[\cite{goles1982}]
  \label{thm:goles3}
  Let $\mathcal{T} = (G,w,b)$ be a threshold 
  network and $x$ an arbitrary configuration. Let us denote by $(x^t)_{t\geq0}$  the orbit of $x$. Then it holds that $L(x^{t+1}) \leq L(x^t)$, for every~$t\geq 0$.
\end{proposition}

In other words, \Cref{thm:goles3} states that the energy function does not increase on the orbits of any threshold network. We  define the \emph{energy delta} for each pair of configurations $x$ and $x'$:
\begin{equation}\label{eqn:delta}
  \Delta L(x',x) = L(x') - L(x) = \sum_{i=1}^{n} \delta_i - 
  \frac{1}{2} (x' - x)^T W(x'-x)\text{.}
\end{equation}
where $\delta_i = - (x'_i - x_i)(\sum_{i=1}^{n} w_{ij} x_j - 
b_i)$.  The proof of \Cref{thm:goles3} uses combinatorial tools to show that  $\Delta L(x^{t+1},x^t)\leq 0$ for every $t\geq 0$. This has an important consequence on the dynamics, limiting the long-term behavior of the system to only fixed points and limit two-cycles under the parallel update scheme. In~\cite{goles2015}, the authors explore under which conditions limit two-cycles may appear. More precisely, they explore a sufficient and necessary conditions on the interaction graph of $\mathcal{T}$ in which the latter situation holds. In this section, we present a generalization of their result to signed graphs. To do so, we start by stating a technical lemma.

\begin{lemma}
  \label{lemma:energy}
  Let $\mathcal{T} =(G,w,b)$ be a threshold 
  network, and let $x$ and $x'$ two configurations such that $x_i \neq x'_i$ for all $i\in[n]$. Then   $\Delta L(x',x) \leq 2 \s(G'(x))$. 
\end{lemma}

\begin{proof} Let $x$ be a configuration defining a two cycle, i.e., if $x' = F(x)$ then $x_i \neq  x'_i$ for all $i\in [n]$.   Using \Cref{eqn:delta} we have that $$\Delta L(x',x) \leq \sum_{i = 1}^{n} \delta_i - \frac{1}{2} 
  (x'-x)^T W (x'-x),$$ where $$\delta_i = - (x'_i - x_i)(\sum_{i=1}^{n} w_{ij} x_j - b_i).$$ Observe that, since $x_i \neq x_i'$ we have that $\delta_i \leq -2$. Indeed, on the one hand,  $(x'_i - x') \in \{-2,2\}$.  On the other hand $(\sum_{i=1}^{n} w_{ij} x_j - b_i) \neq 0$ and has the same sign than$(x'_i - x')$. 
  
  Thus, $$\Delta L(x',x) \leq -2n - \frac{1}{2} (x'-x)^T W (x'-x).$$ 
  
  Now let us define $y = \dfrac{1}{2} (x' - x)\in \{-1,1\}^n$. Then 
  \begin{align*}
      \dfrac{1}{2}(x'-x)^T W (x'-x) &= 2y^TWy = 2 \sum_{i,j} w_{ij}y_iy_j\\
      & = 2\left(\sum_{i \in [n]}\sum_{j\neq i}  w_{ij}y_iy_j + \sum_{i} w_{ii} y_i^2 \right).
  \end{align*}
  We now use that $y_i^2 = 1$ to obtain:
  \begin{align*}
      \dfrac{1}{2}(x'-x)^T W (x'-x) 
      & = 2\sum_{i \in [n]}\sum_{j\neq i} w_{ij} y_iy_j + 2\sum_{i} w_{ii} \\
      &= 4\varphi(y) + 2d^+ - 2d^-.
  \end{align*}

Using \Cref{lem:varphi}  we conclude that
\begin{align*}
\Delta L(x',x) &\leq 2\left(-n -d^+ + d^- -2\varphi(y)\right)\\
&\leq 2\left(-n -d^+ + d^- -2\left(\min_{y\in \{-1,1\}^n} \varphi(y)\right)\right) = 2\s(G,w).
\end{align*}\end{proof}
We are now ready to prove the main theorem of this section.

\begin{theorem}\label{theo:sufficient}
  Let $(G,w)$ be a signed graph satisfying that all its induced signed subgraphs $(G',w)$  satisfy $\s(G',w) < 0$. Then, for every threshold vector $b$, the threshold network $\mathcal{T}=(G,w,b)$ is stable.
\end{theorem}

\begin{proof}
  Observe that, by \Cref{thm:goles1} 
  the signed threshold networks can only admit limit cycles of period $2$ and fixed points. 
  For the sake of contradiction, let us assume that the conditions of the theorem holds, but that there is a threshold vector such that  $\mathcal{T} = (G,w,b)$ admits a limit two-cycle. Let $x$ be a configuration such that $x' = F(x)$ satisfies $x' \neq x$ and $F(x') = x$. Let us define $V^{'} = \{i \in V: x_i \neq x'_i\}$ and the signed subgraph
   $(G',w)$ of $(G,w)$ induced by $V'$.   From \Cref{lemma:energy}, we have that on $(G',w)$, the energy delta is $\Delta L(x',x) \leq 2\s(G') < 0$, and thus $L(x') < L(x)$. 
  But from \Cref{thm:goles3} $L$ is non-increasing, thus we have that $$L(x) = L(F(x')) \leq L(x')<L(x),$$ which is a contradiction. 
\end{proof}

\subsection{Necessary conditions for stability} 

In this subsection we show that the sufficient condition for stability given in \Cref{theo:sufficient} is also necessary. More precisely, we show the following theorem. 

\begin{theorem}\label{theo:necessary}
Let $(G,w)$ be a signed graph containing an induced sugraph $(G',w)$ such that $\s(G',w)\geq 0$. Then, there exists a threshold vector $b$ such that the threshold network $\mathcal{T} = (G,w,b)$ is unstable. 
\end{theorem}

The proof follows from a series of lemmas. The first one states that every total limit cycle of a subset of nodes of a threshold network can be extended to a limit two-cycle network. To do so, we introduce the following notation. Let $\mathcal{T}=(G,w,b)$ be a threshold network and $V'$ be a set of vertices of $G$. The threshold network $\mathcal{T}[V'] = (G[V'], w|_{V'}, b|_{V'})$ is called an \emph{induced threshold sub-network} of $\mathcal{T}$. %Furthermore, we abuse notation and represent $\mathcal{T}[V']$ by $(G[V'], w, b)$. 

\begin{lemma}\label{lem:extension}
Let $\mathcal{T}=(G,w,b)$ be a threshold network and let $V'$ be a set of vertices such that $\mathcal{T}[V']$ is unstable. Then there exists a threshold vector $\hat{b}$ such that $\widehat{\mathcal{T}} = (G,w,\hat{b})$ is unstable. 
\end{lemma}

\begin{proof}
Let $x'$ be a configuration of $V'$ which leads  in $\mathcal{T}[V']$ to an attractor that is not a fixed point. Let us denote $f'_i$ the local function of node $i\in V'$. We extend $x'$ into a configuration $y$ of $V$ by assigning $y_i = -1$ to all nodes $i \in V\setminus V'$.  Then, we fix the state~$-1$ for the nodes in $i \in V\setminus V'$ on the dynamics of $\widehat{\mathcal{T}}$ by assigning to them a threshold $\hat{b}_i = 2n$. Finally, for each $i\in V'$, we define  $$\hat{b}_i = b_i - \sum_{j\in V\setminus V'} w_{ij}. $$
Let $\widehat{F} = (\hat{f}_1, \dots, \hat{f}_n)$ be the global function defined by $\widehat{T}$. We have that for each $i\in V\setminus V'$:
$$\sum_{j \in V} w_{ij}y_j - \hat{b}_i = \sum_{j \in V} w_{ij}y_j - 2n \leq -n <0 \Rightarrow \hat{f}_i(y) = -1 = y_i.$$ 
Furthermore, for each~$i\in V'$:
\begin{align*}
\sum_{j \in V} w_{ij}x_j - \hat{b}_i 
& = \sum_{j\in V'} w_{ij}y_j + \sum_{j\in V\setminus V'} w_{ij}y_j - \left(b_i -  \sum_{j\in V\setminus V'} w_{ij}\right)\\
& = \sum_{j\in V'} w_{ij}x'_j - \sum_{j\in V\setminus V'} w_{ij} - b_i +  \sum_{j\in V\setminus V'} w_{ij}\\
& = \sum_{j\in V'} w_{ij}x'_j - b_i. 
\end{align*}
We obtain that $\hat{f}_i(y) = f'_i(x') = -x'_i = -y_i$. 
We deduce that $y$ defines a limit two-cycle for $\widehat{T}$.
\end{proof}

\begin{remark}
\Cref{lem:extension} implies that the mere existence of a node with a negative self-loop in a signed network is a sufficient condition for instability. Indeed, the threshold network defined by a single node with a negative loop an threshold $0$ has only one attractor consisting on the node alternating between $-1$ and $+1$.  \Cref{lem:extension} states that this threshold subnetwork can be extended to obtain one for the whole network with a limit two-cycle.
\end{remark}

We are now ready to prove \Cref{theo:necessary}.

\begin{proof}[Proof of \Cref{theo:necessary}] Without loss of generality, we can assume that $\s(G,w)\geq 0$ and that $\s(G',w)<0$ for every induced signed subgraph $(G',w)$ of $(G,w)$. Otherwise, we can pick any minimal by inclusion induced subgraph $(G',w)$ of $(G,w)$ such that $\s(G',w)\geq 0$, and prove the existence of an unstable threshold network defined over $(G',w)$. Using \Cref{lem:extension} we then extend the obtained unstable threshold network to one defined over $(G,w)$. 

Now, we start using \Cref{lem:varphi} to write
$$\s(G,w) =  -n -d^+ + d^- - 2\left(\min_{y\in\{-1,1\}^n}\varphi(y)\right),$$
and we pick a configuration $x\in \{-1,1\}^n$ that reaches the minimum of $\varphi(x)$. Then,
\begin{align*}
0\leq \s(G,w) 
&=  -n -d^+ + d^- - 2\varphi(x)\\
&= -n -d^+ + d^- - \left(\sum_{i \in [n]}\sum_{j\neq i} w_{ij}x_ix_j \right)\\
&= \sum_{i\in [n]} \left(-1- w_{ii} - \sum_{j\neq i} w_{ij}x_ix_j   \right). 
\end{align*}
Let $V'\neq \emptyset$ be the largest set of nodes such that, for each $i\in V'$,
$$ \sum_{j\neq i} w_{ij}x_ix_j + w_{ii} + 1  \leq  0.$$
Observe that if $V'=V$, we can use \Cref{lemma:psws} to show that $\mathcal{T}=(G,w,0)$ admits a total limit cycle, and in particular $\mathcal{T}$ is unstable. 

In the following, we show that $V\neq V'$ contradicts the minimality of $(G,w)$. More precisely, suppose that there is a $k\in V\setminus V'$, and denote by $G-k$ the graph induced by $V\setminus\{k\}$. In the following, we show that $\s(G - k, w) \geq \s(G,k)$, contradicting our first assumption. 

Observe first that $k\notin V'$, then
\begin{equation}\label{eq:hmm}
    \sum_{j\neq k} w_{kj}x_kx_j  > -w_{kk}-1.
\end{equation}
We now distinguish two cases. \\

\noindent{\bf Case 1: $w_{kk}=0$.}  In this case we have that \Cref{eq:hmm} becomes
$$ \sum_{j\neq k} w_{kj}x_kx_j  > -1 \Rightarrow \sum_{j\neq k} w_{kj}x_kx_j  \geq 0. $$
If we define $\displaystyle{\varphi'(y) = \dfrac{1}{2}\sum_{i\in [n]\setminus \{k\}}\sum_{j\neq i,j\neq k} w_{ij}y_iy_j}$, we obtain that
\begin{align*}
\s(G-k,w) &= -(n-1) - d^+ + d^-  -2\left( \min_{y\in \{-1,1\}^{n-1}} \varphi'(y)\right).
\end{align*}
Furthermore, $\varphi(x) \geq \varphi'(x)$. Indeed,
\begin{align*}
    \varphi(x) &= \dfrac{1}{2} \sum_{i\in [n]}\sum_{j\neq i} w_{ij}x_ix_j\\
    &= \dfrac{1}{2} \sum_{i\in [n]\setminus {k}}\sum_{j\neq i, j\neq k} w_{ij}x_ix_j + \sum_{j\neq k} w_{kj}x_kx_j\\
    &\geq \dfrac{1}{2} \sum_{i\in [n]\setminus {k}}\sum_{j\neq i, j\neq k} w_{ij}x_ix_j \\
    &= \varphi'(x).
\end{align*}
Hence,\begin{align*}
    \s(G,w) &= -n -d^+ + d^- -2\varphi(x)\\
    &\leq -n -d^+ + d^- -2\varphi'(x)\\
    &\leq \s(G-k,w).
\end{align*} 

\noindent{\bf Case 2: $w_{kk}=1$.}  In this case \Cref{eq:hmm} becomes
$$\sum_{j\neq k} w_{kj}x_kx_j  > -2 ~\Rightarrow~ \sum_{j\neq k} w_{kj}x_kx_j  \geq -1.$$

Remember the definition of the stability index:
$$\s(G,w) = -n - d^+ + d^- +2m - 4\rho(G,w),$$
which applied to $(G-k,w)$ gives 
\begin{align*}
\s(G-k,w) &= -(n-1) - (d^{+}-1) + d^{-} + 2(m - |N(k)|) - 4\rho(G-k,w)\\
& = \s(G,w) - 2(|N(k)|-1) + 4(\rho(G,w) -\rho(G-k,w)).
\end{align*}

Now, for each $a\in \{-1,1\}$ let us define the sets:
$$U^{a}(x,k) = \{j \in N(k): x_j = x_k, w_{kj} = a\},$$
$$D^{a}(x,k) = \{j \in N(k): x_j \neq x_k, w_{kj} = a\},$$
which leads us to obtain that for each $k\in [n]$,
$$\sum_{j\neq i} w_{kj}x_kx_j  = |U^1(x,k) \cup D^{-1}(x,k)| - |U^{-1}(x,k) \cup D^1(x,k)| \geq -1,$$
and thus that
$$ |U^1(x,k) \cup D^{-1}(x,k)| +1 \geq  |U^{-1}(x,k) \cup D^1(x,k)|.$$
Observe that  each $j\in U^{a}(x,k)$ satisfies that  $\{k,j\}\in U^{1,a}(x) \cup U^{-1,a}$, and  each $j\in D^{a}(x,k)$ satisfies $\{k,j\} \in D^{a}(x)$. Then, 
\begin{align*}
    \rho(G-k,w) &\leq  |U^{1,1}(x)\cup U^{-1,1}(x)\cup D^{-1}(x)| - |U^1(x,k) \cup D^{-1}(x,k)|\\
    &= \rho(G,w) - |U^1(x,k) \cup D^{-1}(x,k)|.
\end{align*}
Since the sets $U^{1}(x,k)$, $U^{-1}(x,k)$, $D^1(x,k)$ and $D^{-1}(x,k)$ form a partition of $N(k)$,  we have that $$|U^1(x,k) \cup D^{-1}(x,k)| \geq   \dfrac{|N(k)|-1}{2}.$$

 We deduce that $$\rho(G,w) - \rho(G-k,w) \geq |U^1(x,k) \cup D^{-1}(x,k)| \geq \dfrac{|N(k)|-1}{2}$$ and obtain
\begin{align*}
    \s(G,w-k) &= \s(G,w) - 2(|(|N(k)|-1) + 4(\rho(G,w)-\rho(G,w-k))\\ 
    &\geq
    \s(G,w) - 2(|N(k)|-1) + 4 \left(\dfrac{|N(k)|-1}{2}\right)\\
    &\geq \s(G,w).
\end{align*}
In any case we prove the existence of a node $k\in[n]$ such that $\s(G-k,w) \geq \s(G,w)$, contradicting our first assumption. We deduce that $V' = V$ and that $\mathcal{T} = (G,w,0)$ is unstable.   
\end{proof}

\section{Periodic dynamics}
\label{sec:periodic}
In this section we study the dynamics from a more general standpoint, by focusing 
on periodic update schemes. We show notably that with no assumptions, long cycles (i.e., super polynomial cycles in the size of the network) may appear. Remember that a periodic update scheme is a sequence $\mu = (I_1, \ldots, I_{\ell})$ such that $I_{k} \in \mathcal{P}(V)$, where $\mathcal{P}(V)$ is the power set of $V$.  From now on, for a periodic update scheme $\mu = (I_1, \ldots, I_{\ell})$, we call $G(I_{k})$ the subgraph induced by the set of nodes $I_{k}$.

\subsection{A sufficient condition for stability}

The following theorem is an extension of Theorem~\ref{theo:sufficient} to the 
 family of periodic update schemes. 

\begin{theorem}\label{teo:period}
  Let $(G,w)$ be signed graph and let us consider a periodic update scheme $\mu = (I_1, \ldots, 
  I_p)$. 
  If for all $1 \leq k \leq p$, for all the induced signed subgraphs $(G',w)$ of $(G[I_k],w)$ satisfy $\s(G',w) < 0$ then, for every threshold vector $b,$ the threshold network $\mathcal{T}=(G,w,b,\mu)$  is stable.   
\end{theorem}

We start by showing a technical lemma:

\begin{lemma}\label{lemma:endec}
    Let $(G,w)$ be a signed graph and $\mu$ be a periodic update scheme of period $p.$ Let $x$ be a configuration and  $k \leq p$. Let us call $I = \mu(k)$. If the $f_{I}(x) \not = x$ then,   $$\Delta L(x',x) \leq 2\s(G[I'],w),$$ where $x' = f_I(x)$ and $I' = \{i \in I: x_i \not = x'_i\}.$
\end{lemma}

\begin{proof}
    Let us fix a periodic update scheme $\mu$ with some period $p$ and let us assume that $f_I(x)_i \not = x_i$ for some configuration $x$ and for all $i \in I,$  where $k \leq p$ such that $\mu(k) = I.$ Let us call $x' = f_I(x).$ If we proceed in the same way that the proof of Lemma \ref{lemma:energy} and we use Equation \ref{eqn:delta}, we have that: 
$$\Delta L(x',x) \leq -2 |I| - 2\left(\sum_{i \in I}\sum_{j \in I, j\neq i}  w_{ij}y_iy_j + \sum_{i \in I} w_{ii} y_i^2 \right),$$

where $y=x'-x.$ Thus, we deduce that:

$$\Delta L(x',x) \leq 2\s(G[I],w).$$
Now,  observe that if $I$ is such that $f_I(x) \not = x$ and we call $x' = f_I(x),$ we can proceed as same as in the proof of Theorem \ref{theo:sufficient}, and work with the induced signed subgraph $(G[I'],w)$ where the set $I'$ is defined as $I' = \{i \in I: x_i \not = x'_i\}.$ In fact, in that case we have:
$$\Delta L(x',x) \leq 2\s(G[I'],w).$$
\end{proof}

Now we proceed with the proof of~\Cref{teo:period}.
\begin{proof}[Proof of \Cref{teo:period}]
Observe that for each $k \in \{0, 
\ldots, \ell\}$, the nodes that may change state are the nodes in $I_k$ 
whereas all the other remains in their current state.  

 Now, let us assume that $x$ is a configuration such that $F_\mu(x) \not = x$. Let us define the sequence $x_1, \hdots, x_p$ as $x_{q} = f_{I_q}(x_{q-1})$  for $1 \leq q \leq p,$ where we consider that $x_{1} = x$ and we have that $x_{p} = x'.$  Observe that, since $F_\mu(x) \not = x,$ there must exists some $1\leq k\leq p$ such that $x_k \not = x_{k-1}.$ 

Let us call $r_1, \hdots, r_{\ell}$ be the collection of time-steps such that $x_{r_{q}} \not =x_{r_{q}-1}$ for $1\leq q \leq \ell.$ Since we have that:

$$\Delta L (x_p,x_1) = L(x_{p}) - L(x_1) = \sum \limits_{q=1}^{p} L(x_{q+1})-L(x_{q}), $$
and that $L(x',x) = 0$ whenever $x'=x,$ we can assume that $r_1\hdots r_{\ell}$ are consecutive time steps and that $r_q = q$ for $1 \leq q \leq \ell.$

Now, we compute the energy difference between a configuration $x$ and $x' = F_{\mu}(x)$ and we use Lemma \ref{lemma:endec}:
$$\Delta L (x_p,x_1) = L(x_{\ell}) - L(x_1) = \sum \limits_{q=1}^{\ell} L(x_{q+1})-L(x_{q}) \leq 2\sum \limits_{q=1}^{\ell} S(G[I'_q],w)),$$

where $I'_q = \{i \in I_q: x_i \not = x'_i\}.$ Since for all $1 \leq q \leq p$ and for all set $I_{q}$ we have that  $S(G',w) <0,$ for all induced signed subgraph $(G',w)$ of $(G[I_{q}],w),$  we have that 
$$\Delta L (x_p,x_1) = L(x_{\ell}) - L(x_1) = \sum \limits_{q=1}^{\ell} L(x_{q+1})-L(x_{q}) \leq  2\sum\limits_{q=1}^{\ell} S(G[I'_q],w) < 0.$$

Finally, let us consider an attractor $x^1\hdots, x^p$ of period  $p$ for $F_{\mu},$ i.e., $x^{t} = F_{\mu}^{t-1}(x^1)$ for $t\geq 2$ and  $F_{\mu}(x^p) = x^1.$ Let us assume that $p> 1.$ By the previous calculations we have that:

$$L(x^p)<L(x^{p-1})<L(x^{p-2})< \hdots < L(x^2) < L(x^1). $$
In addition, we have that $L(F_{\mu}(x^p)) \leq L(x^p).$ However, by definition, we also have that $F_{\mu}(x^p) = x^1,$ and thus,

$$L(x^1) = L(F_{\mu}(x^p)) \leq L(x^p)<L(x^1),$$

which is a contradiction.
\end{proof}

\begin{remark}
\begin{enumerate}
   \item  If the condition on the subgraphs of the sets $I_q$ on the previous theorem does not hold, it has been shown in \cite[Theorem 4]{donoso2024}, that limit cycles of superpolynomial period on the size of the network may appear. Interestingly, this result holds for very simple graphs such as cycle graphs. More precisely, it is shown that the elementary cellular automaton rule labeled as rule $178$ in Wolfram's notation (which can be seen as a threshold network, more precisely, an unstable majority network) exhibits limit cycles of superpolynomial period on the size of the network for a particular periodic update scheme. Observe that in the case presented in the latter paper, since the underlying graph is a cycle with negative loops, we have that $\s(G,w) = - n - 0 + n + 2n - 4 \times 0 = 2n > 0.$ 
 
\item It is interesting to observe that this latter result somehow generalizes the one shown in \cite{goles2015}, in which the authors also find that, in the unsigned case, for some block sequential update schemes, it is possible to construct threshold networks exhibiting limit cycles of superpolynomial period for some threshold vector. However, in the signed case equipped with a periodic update scheme, the structure of the network is simpler than in the unsigned case.
\end{enumerate}

\end{remark}

\subsection{Stability for update schemes with small blocks.}
Now, we turn our attention to the problem of stability of networks that are updated with a periodic update scheme in which the size of the blocks is of some (constant) size of at most  $h \geq 1.$ This problem was already tackle in \cite{goles2015} for unsigned threshold networks. In that paper, the authors performed an exhaustive study of graphs with size at most $3$ and self-loops in each node. They characterized completely the stability of these graphs. In particular, they have shown (by explicitly computing the stability index for each possible graph) that all the graphs with less than three nodes are stable.  

In addition, the authors of \cite{goles2015} computed all the possible connected graphs with positive self-loops in all the nodes of size $4$ and $5.$ Intertestingly, some of these graphs exhibit total limit cycles. However, there were only two graphs having this behavior. As a consequence, if they are forbidden as induced subgraphs of the original graph then, the stability result can be extended for $h \leq 5.$

As we will see in this section, this result is no longer true for signed graphs. In fact, we can only have the stability property for $h=1$ and $h=2$, i.e. there are some graphs of size $3$ admiting total-two cycles for some threshold vector.

In this section, we show that the stability result holds without any additional assumptions for $h\leq 2$ and then we provide a list of all the graphs for $3\leq h \leq 4$ (with positive loops in all the nodes) admitting total-two cycles (see Table \ref{tab:allg}). As we will see in the next corollaries, if a periodic update scheme $\mu$ is such that no induced subgraph of a block is on the list then, the graph is stable for $\mu$.

Remember that we say that a signed graph $(G,w)$ is \emph{stable} for some periodic update scheme $\mu$ if $\mathcal{T} = (G,w,b,\mu)$ admits only fixed points for all threshold vector $b$. Conversely, $(G,w,\mu)$ is unstable for $\mu$ if there exists a threshold vector $b$ such that $\mathcal{T} = (G,w,b,\mu)$ admits limit cycles of period $q>1.$

First, we show the result for partitions of size $1$ and $2.$

\begin{corollary} \label{cor:partriv}
    Let $(G,w)$ be a signed graph such that $w_{ii} = 1$ for all $i \in V.$ Let $\mu$ be a periodic update scheme of period $p\geq 1$ such that $|\mu(k)| = |I_k|\leq 2,$ for all $1\leq k\leq p.$ Then, $(G,w)$ is stable for $\mu.$ Otherwise, if for some $k$ we have that $|\mu(k)| = 3$ then,  $(G,w)$ is unstable for $\mu$.
\end{corollary}
\begin{proof}
Observe that as a consequence of~\Cref{teo:period} it suffices to show $\s(G)<0$ for graphs $G$ of size $1$ and $2.$ First, observe that if $G$ is such that $|V| = 1$ then $\s(G) = -1-w_{ii}.$  Thus, $\s(G)<0.$ For the case $|\mu(k)| = |I_k| = 2,$ observe that any connected graph $G$ with two nodes such that $w_{ii} = 1$ for all $i \in V.$ satisfies $S(G) = -2 -2 + 0 + 2\times 1 + 0 = -2 <0.$ If for some $k$ it is allowed to have a block of size at least $3$ (then $G$ can have an induced subgraph of size $3$), we can define an example exhibiting two-cycles by considering one of the graphs with three nodes in \Cref{tab:allg} and using \Cref{lem:extension} to extend the total-cycle to an attractor of period $q>1.$ 

\end{proof}
Observe that for blocks of bigger size the result is no longer true. In fact, we provide in \Cref{tab:allg} a list of graphs of size $3$ and $4$ with positive loops in each node (up to isomorphism) admitting total limit cycles. However, as we show in the next corollary, if these graphs are not induced subgraphs of the original network then, the stability result holds for blocks of size at most $4.$
\begin{corollary} \label{cor:par}
    Let $(G,w)$ be a signed graph such that $w_{ii} = 1$ for all $i \in V.$ Let $\mu$ be a periodic update scheme of period $p\geq 1$ such that $|\mu(k)| = |I_k|\leq 4,$ for all $1\leq k\leq p.$ If $G[I_k]$ is not a graph (up to isomorphism) shown in Table \ref{tab:allg} then, $(G,w)$ is stable for $\mu.$  Otherwise, $(G,w)$ is unstable for $\mu.$
\end{corollary}

\begin{proof}
   As a consequence of Theorem \ref{teo:period}, it suffices to show $\s(G)<0$ for graphs with $n$-nodes and $1\leq n \leq 4.$ The case $n = 1$ and $n=2$ has been shown in Corollary \ref{cor:partriv}. For the cases $n=3$ and $n=4$ we provide a list of all the graphs with positive self-loops in all the nodes admiting a total two cycle in Table \ref{tab:allg}. If these cases are excluded (no induced subgraph is on the list provided in latter figure) then, $\s(G)<0.$ If for some $k$ it is allow to have a block of size at least $3$ (then $G$ can have an induced subgraph of size $3$ or $4$) and thus, we can define an example exhibiting two-cycles by considering some graph in Table \ref{tab:allg} and using Lemma \ref{lem:extension} to extended the total-cycle to an attractor of period $q>1.$
\end{proof}

\section{Discussion}

In this paper, we have presented a graph parameter, the stability index, 
which relates the dynamics of a threshold network with the structure of the 
underlying signed graph. 
The sign of this parameter for subgraphs allows us to determine whether the 
dynamics is stable or not. 
However, as mentioned at the beginning of the article, computing this index 
could be very impractical (computing $\rho$ is \textbf{NP}-hard in general). 
A particularly interesting approach would consist in studying 
update schemes induced by sets of bounded size. 
For example,  if one study the family of all update 
schemes with at most $3$ nodes, the stability for connected graphs can be 
characterized in terms of forbidden subgraphs (notably signed triangles). 
An exhaustive study of different subgraphs could be a promising approach. 
Finally, it could be also interesting to study the structure of particular 
graphs that may be of interest of some applications such as 
regulatory networks or social networks.

\clearpage
\begin{table}[ht]
\centering
\resizebox{\textwidth}{!}{%
\begin{tabular}{cccc}
\begin{tabular}{c}\includegraphics[width=0.23\textwidth]{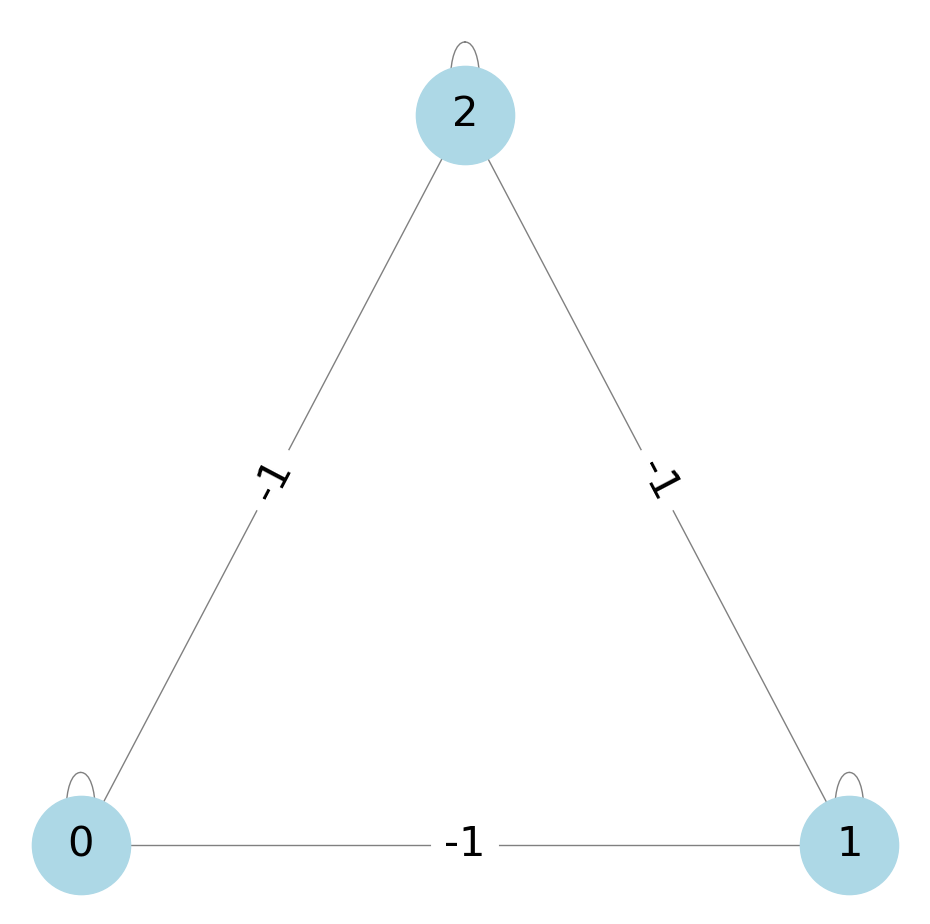} \\ \tiny $\mathcal{S}(G) = 0$ \\ \tiny $\overline{x}$ = \begin{tabular}{|c|c|c|} \hline 0&1&2 \\ \hline -1& -1& -1 \\ \hline\end{tabular} \\ \end{tabular} & 
\begin{tabular}{c}\includegraphics[width=0.23\textwidth]{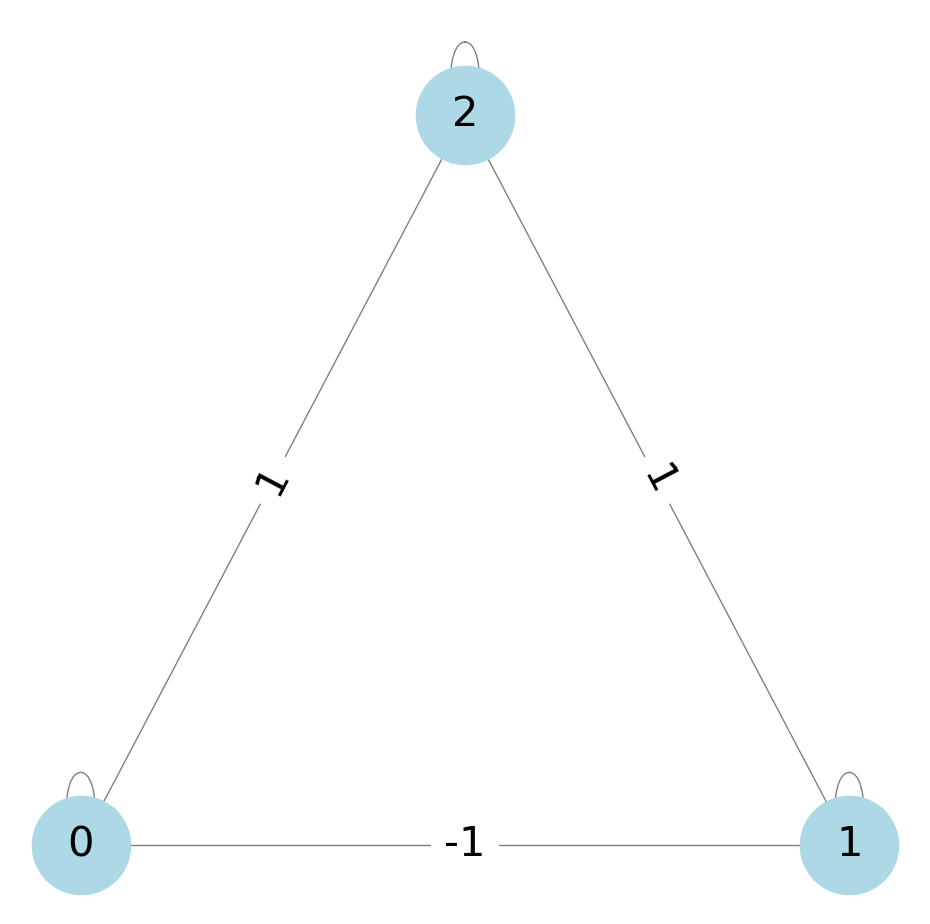} \\ \tiny $\mathcal{S}(G) = 0$ \\ \tiny $\overline{x}$ = \begin{tabular}{|c|c|c|} \hline 0&1&2 \\ \hline -1& -1& 1 \\ \hline\end{tabular} \\ \end{tabular} & 
\begin{tabular}{c}\includegraphics[width=0.23\textwidth]{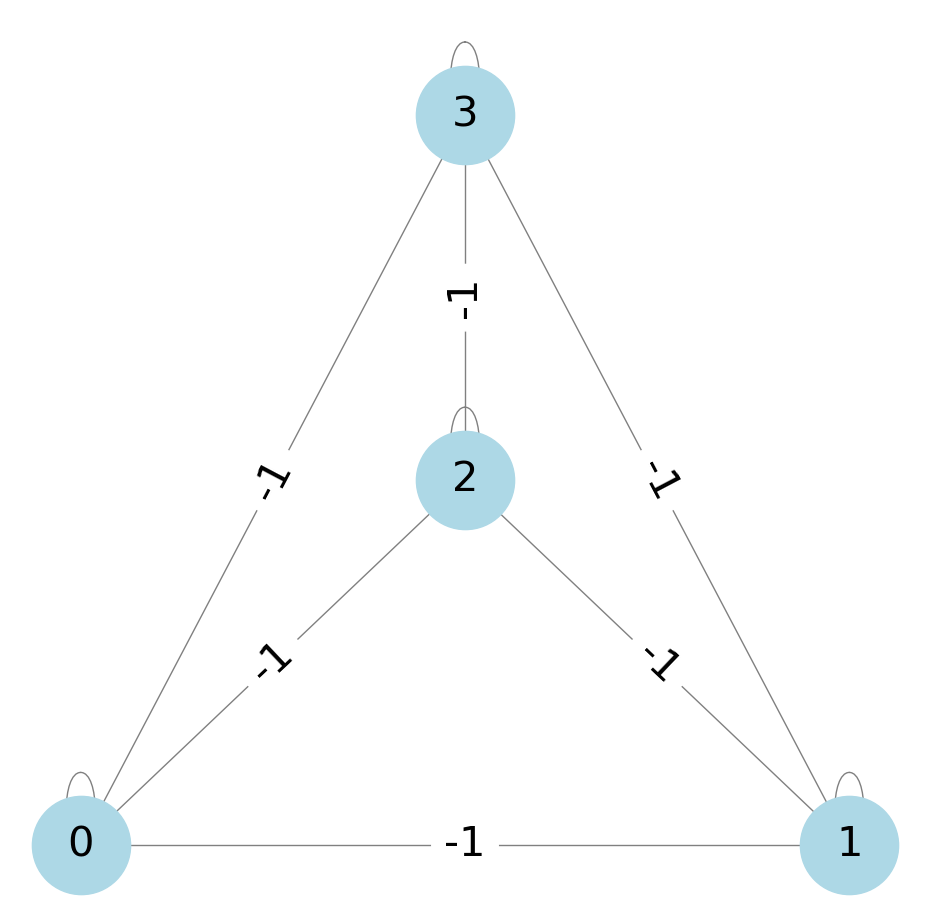} \\ \tiny $\mathcal{S}(G) = 4$ \\ \tiny $\overline{x}$ = \begin{tabular}{|c|c|c|c|} \hline 0&1&2&3 \\ \hline -1& -1& -1& -1 \\ \hline\end{tabular} \\ \end{tabular} & 
\begin{tabular}{c}\includegraphics[width=0.23\textwidth]{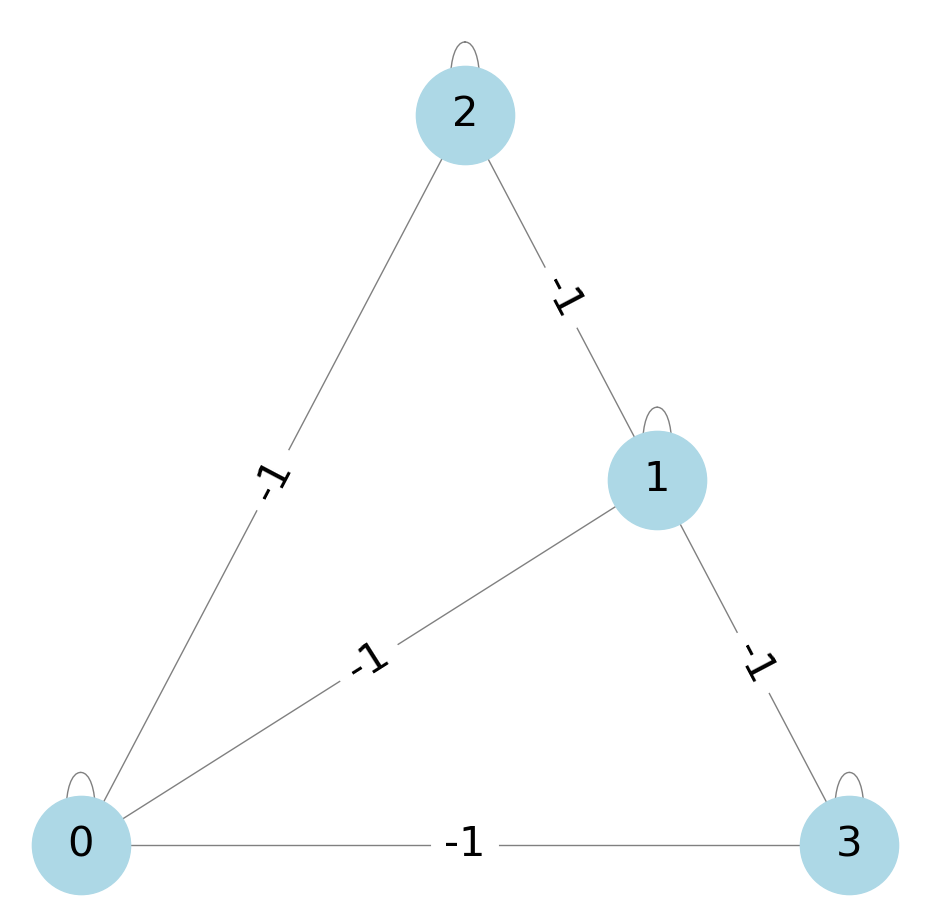} \\ \tiny $\mathcal{S}(G) = 2$ \\ \tiny $\overline{x}$ = \begin{tabular}{|c|c|c|c|} \hline 0&1&2&3 \\ \hline -1& -1& -1& -1 \\ \hline\end{tabular} \\ \end{tabular} \\
\begin{tabular}{c}\includegraphics[width=0.23\textwidth]{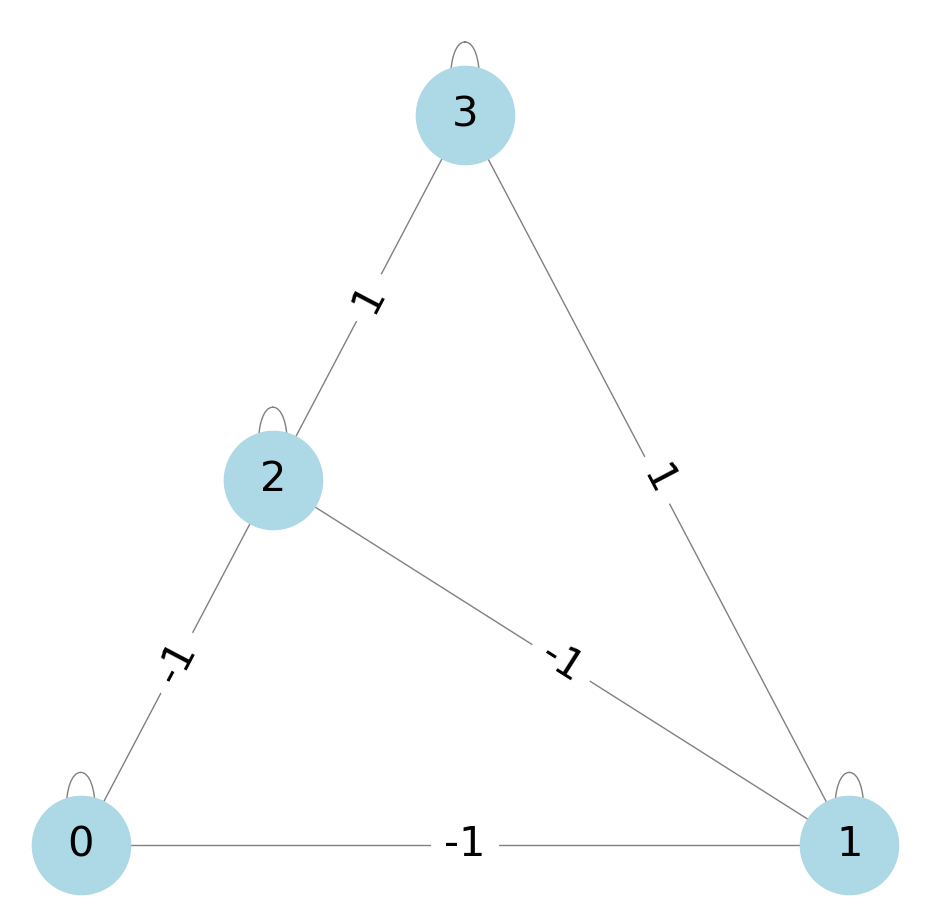} \\ \tiny $\mathcal{S}(G) = 2$ \\ \tiny $\overline{x}$ = \begin{tabular}{|c|c|c|c|} \hline 0&1&2&3 \\ \hline -1& -1& -1& 1 \\ \hline\end{tabular} \\ \end{tabular} & 
\begin{tabular}{c}\includegraphics[width=0.23\textwidth]{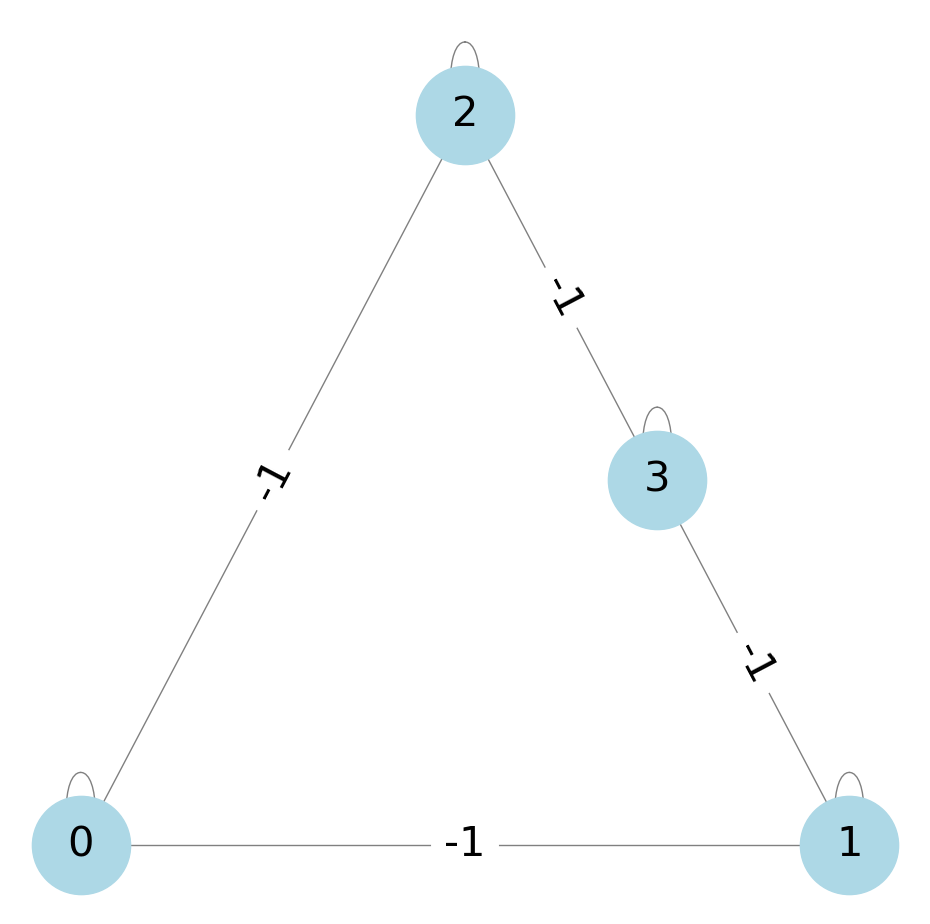} \\ \tiny $\mathcal{S}(G) = 0$ \\ \tiny $\overline{x}$ = \begin{tabular}{|c|c|c|c|} \hline 0&1&2&3 \\ \hline -1& -1& -1& -1 \\ \hline\end{tabular} \\ \end{tabular} & 
\begin{tabular}{c}\includegraphics[width=0.23\textwidth]{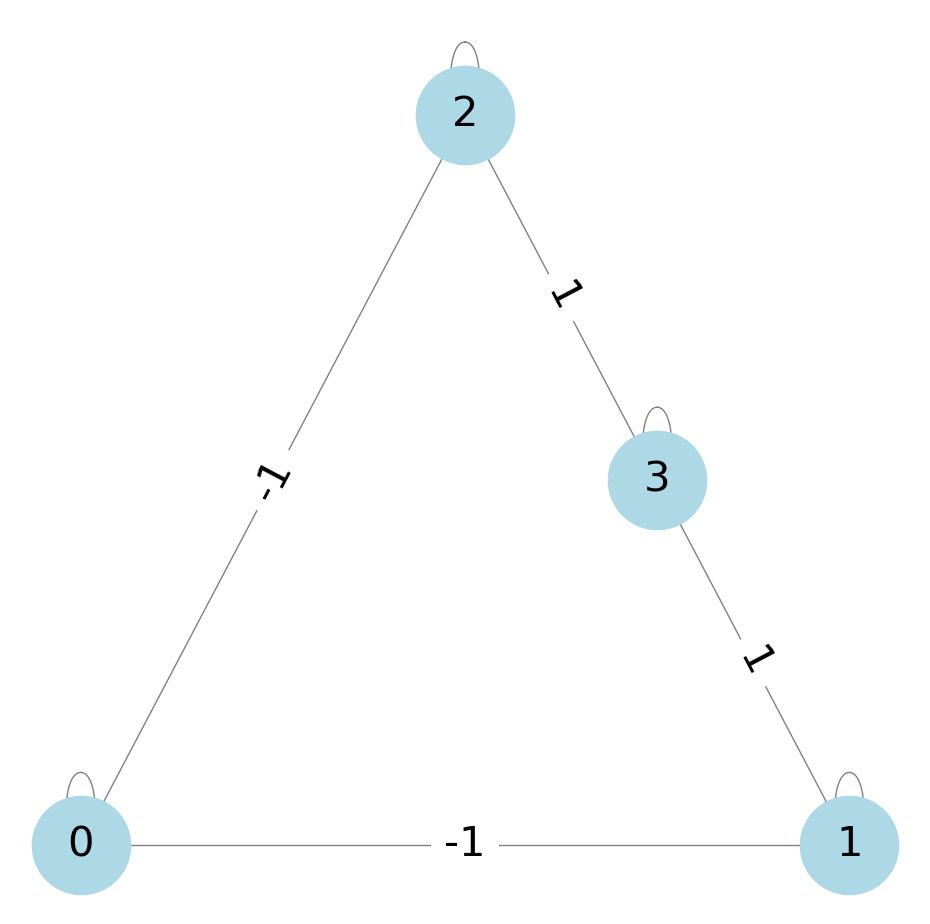} \\ \tiny $\mathcal{S}(G) = 0$ \\ \tiny $\overline{x}$ = \begin{tabular}{|c|c|c|c|} \hline 0&1&2&3 \\ \hline -1& -1& -1& 1 \\ \hline\end{tabular} \\ \end{tabular} & 
\begin{tabular}{c}\includegraphics[width=0.23\textwidth]{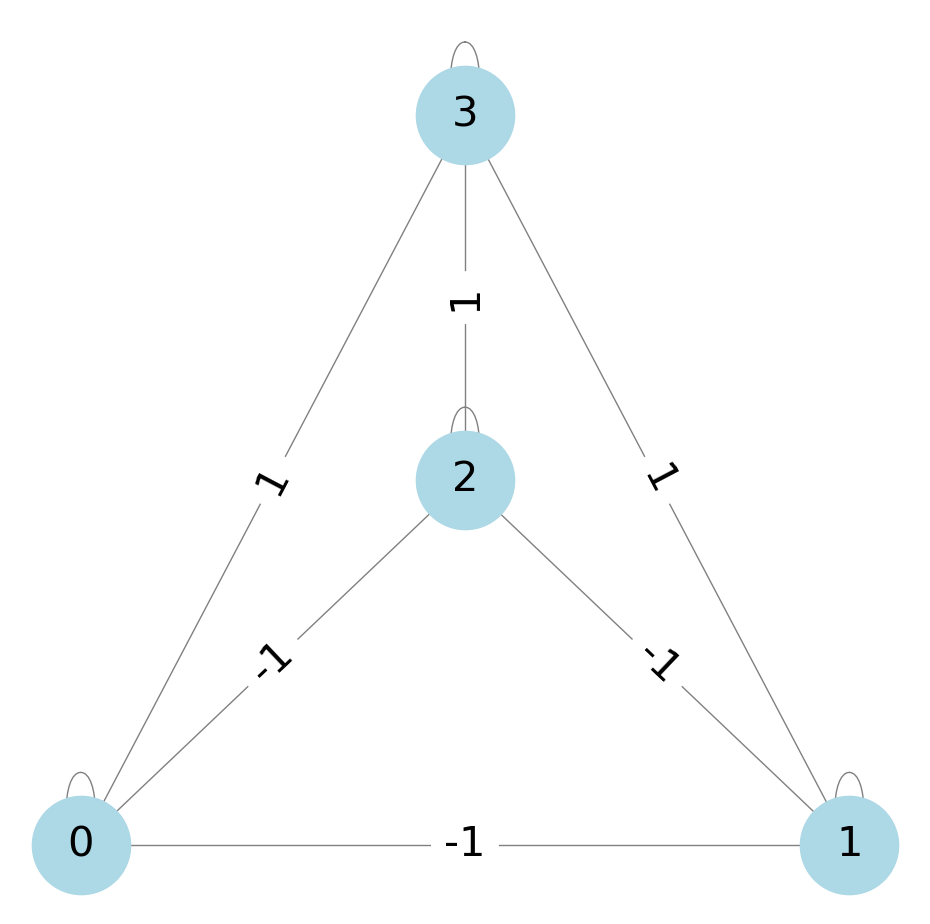} \\ \tiny $\mathcal{S}(G) = 4$ \\ \tiny $\overline{x}$ = \begin{tabular}{|c|c|c|c|} \hline 0&1&2&3 \\ \hline -1& -1& -1& 1 \\ \hline\end{tabular} \\ \end{tabular} \\
\begin{tabular}{c}\includegraphics[width=0.23\textwidth]{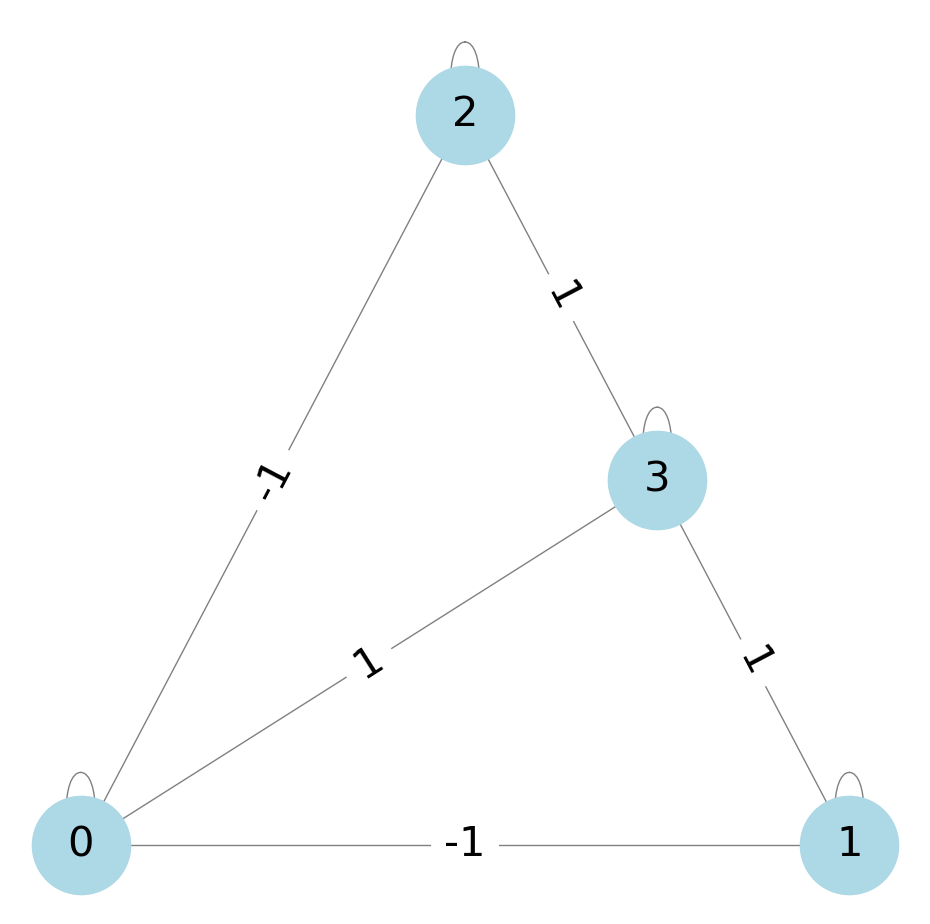} \\ \tiny $\mathcal{S}(G) = 2$ \\ \tiny $\overline{x}$ = \begin{tabular}{|c|c|c|c|} \hline 0&1&2&3 \\ \hline -1& -1& -1& 1 \\ \hline\end{tabular} \\ \end{tabular} & 
\begin{tabular}{c}\includegraphics[width=0.23\textwidth]{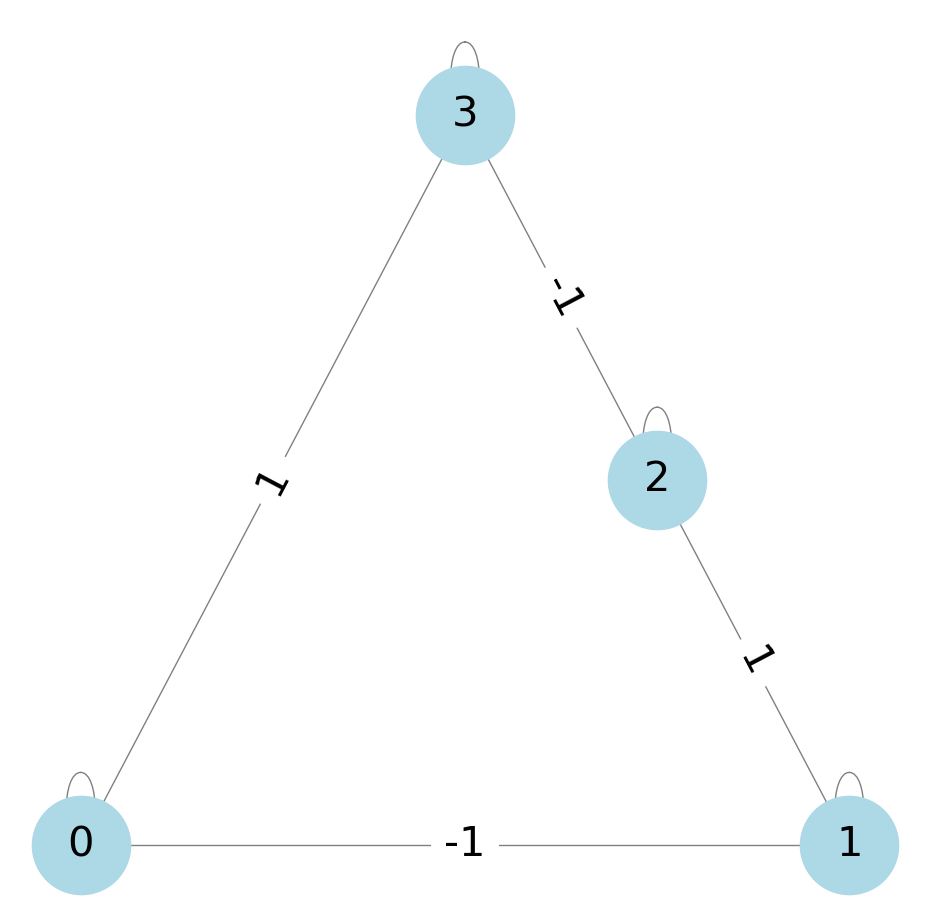} \\ \tiny $\mathcal{S}(G) = 0$ \\ \tiny $\overline{x}$ = \begin{tabular}{|c|c|c|c|} \hline 0&1&2&3 \\ \hline -1& -1& 1& 1 \\ \hline\end{tabular} \\ \end{tabular} & 
\begin{tabular}{c}\includegraphics[width=0.23\textwidth]{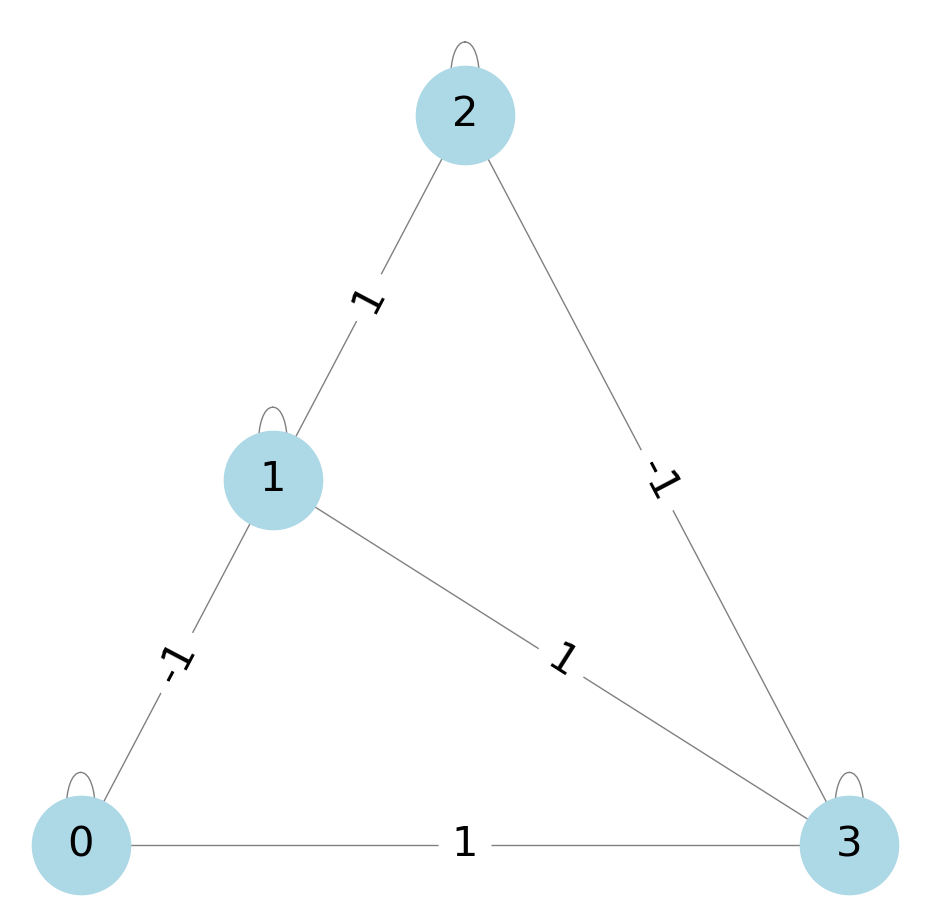} \\ \tiny $\mathcal{S}(G) = 2$ \\ \tiny $\overline{x}$ = \begin{tabular}{|c|c|c|c|} \hline 0&1&2&3 \\ \hline -1& -1& 1& 1 \\ \hline\end{tabular} \\ \end{tabular} & 
\begin{tabular}{c}\includegraphics[width=0.23\textwidth]{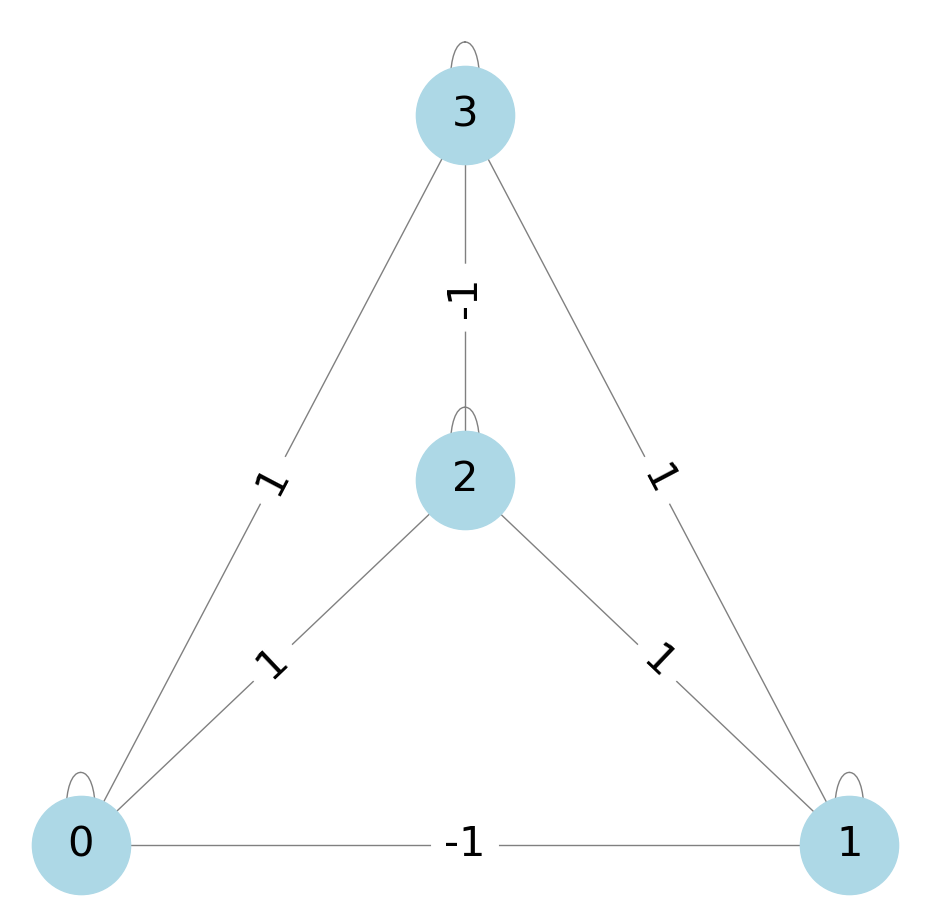} \\ \tiny $\mathcal{S}(G) = 4$ \\ \tiny $\overline{x}$ = \begin{tabular}{|c|c|c|c|} \hline 0&1&2&3 \\ \hline -1& -1& 1& 1 \\ \hline\end{tabular} \\ \end{tabular} \\
\begin{tabular}{c}\includegraphics[width=0.23\textwidth]{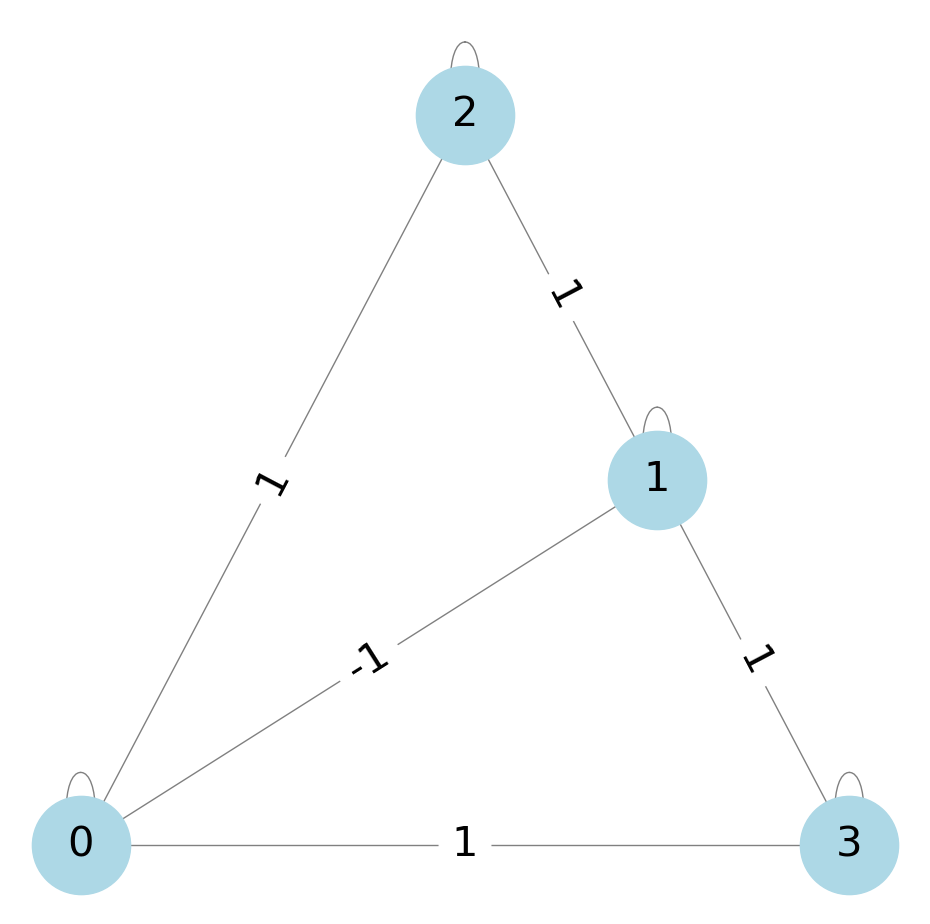} \\ \tiny $\mathcal{S}(G) = 2$ \\ \tiny $\overline{x}$ = \begin{tabular}{|c|c|c|c|} \hline 0&1&2&3 \\ \hline -1& -1& 1& 1 \\ \hline\end{tabular} \\ \end{tabular} & 
\begin{tabular}{c}\includegraphics[width=0.23\textwidth]{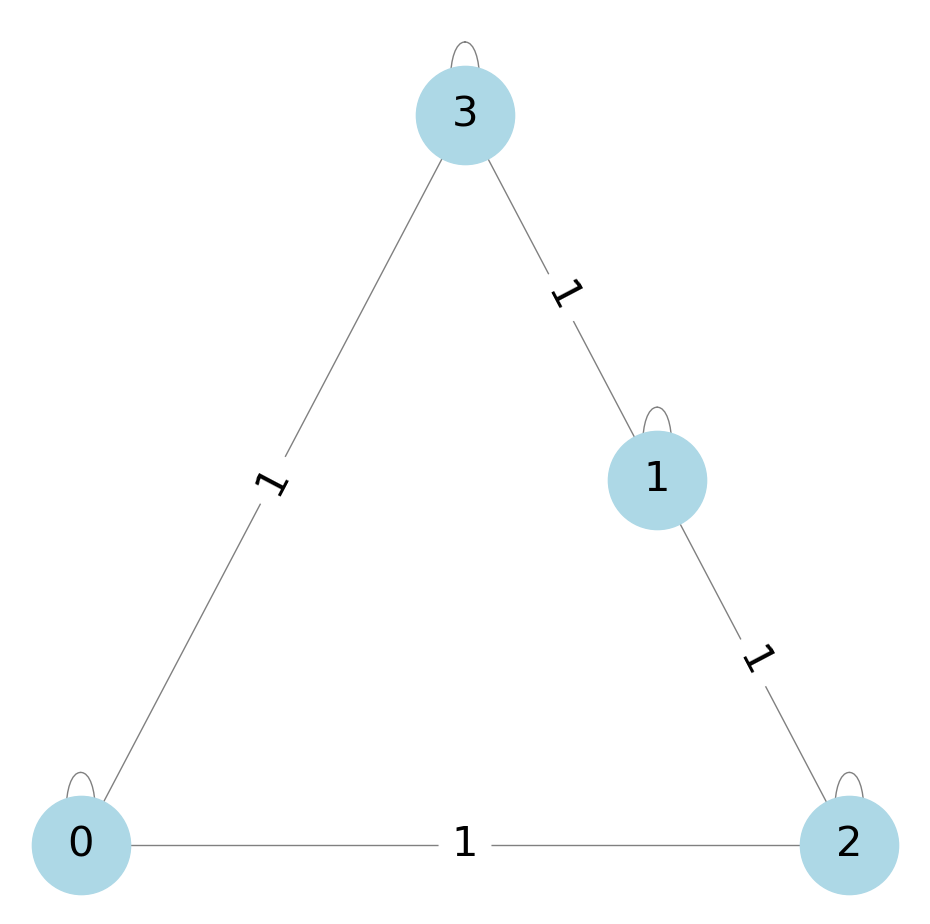} \\ \tiny $\mathcal{S}(G) = 0$ \\ \tiny $\overline{x}$ = \begin{tabular}{|c|c|c|c|} \hline 0&1&2&3 \\ \hline -1& -1& 1& 1 \\ \hline\end{tabular} \\ \end{tabular} & & \\
\end{tabular}
}
\caption{All the signed graphs with $3 \leq n \leq 4$ admitting total-two cycles (up to isomorphism). In the second and third row of each cell we show the value of the stability index of the graph and  the states of a configuration in a total-two cycle ($\overline{x}$). The first row of $\overline{x}$ is an enumeration of the nodes in the graph and the second row is the actual state of each node.}
\label{tab:allg}
\end{table}

\clearpage

\bibliography{biblio.bib}   % Bibliografia

\end{document}